\PassOptionsToPackage{unicode}{hyperref}
\PassOptionsToPackage{hyphens}{url}
\PassOptionsToPackage{dvipsnames,svgnames,x11names}{xcolor}
\documentclass[
]{article}
\usepackage{amsmath,amssymb}
\usepackage{lmodern}
\usepackage{iftex}
\ifPDFTeX
  \usepackage[T1]{fontenc}
  \usepackage[utf8]{inputenc}
  \usepackage{textcomp} 
\else 
  \usepackage{unicode-math}
  \defaultfontfeatures{Scale=MatchLowercase}
  \defaultfontfeatures[\rmfamily]{Ligatures=TeX,Scale=1}
\fi
\IfFileExists{upquote.sty}{\usepackage{upquote}}{}
\IfFileExists{microtype.sty}{
  \usepackage[]{microtype}
  \UseMicrotypeSet[protrusion]{basicmath} 
}{}
\makeatletter
\@ifundefined{KOMAClassName}{
  \IfFileExists{parskip.sty}{%
    \usepackage{parskip}
  }{
    \setlength{\parindent}{0pt}
    \setlength{\parskip}{6pt plus 2pt minus 1pt}}
}{
  \KOMAoptions{parskip=half}}
\makeatother
\usepackage{xcolor}
\usepackage[margin=1in]{geometry}
\usepackage{color}
\usepackage{fancyvrb}

\DefineVerbatimEnvironment{Highlighting}{Verbatim}{commandchars=\\\{\}}
\usepackage{framed}
\definecolor{shadecolor}{RGB}{248,248,248}
\newenvironment{Shaded}{\begin{snugshade}}{\end{snugshade}}

\newcommand{\AttributeTok}[1]{\textcolor[rgb]{0.77,0.63,0.00}{#1}}

\newcommand{\CommentTok}[1]{\textcolor[rgb]{0.56,0.35,0.01}{\textit{#1}}}

\newcommand{\ConstantTok}[1]{\textcolor[rgb]{0.00,0.00,0.00}{#1}}
\newcommand{\ControlFlowTok}[1]{\textcolor[rgb]{0.13,0.29,0.53}{\textbf{#1}}}

\newcommand{\DecValTok}[1]{\textcolor[rgb]{0.00,0.00,0.81}{#1}}

\newcommand{\FloatTok}[1]{\textcolor[rgb]{0.00,0.00,0.81}{#1}}
\newcommand{\FunctionTok}[1]{\textcolor[rgb]{0.00,0.00,0.00}{#1}}

\newcommand{\NormalTok}[1]{#1}

\newcommand{\OtherTok}[1]{\textcolor[rgb]{0.56,0.35,0.01}{#1}}

\newcommand{\SpecialCharTok}[1]{\textcolor[rgb]{0.00,0.00,0.00}{#1}}

\usepackage{longtable,booktabs,array}
\usepackage{calc} 
\usepackage{etoolbox}
\makeatletter
\patchcmd\longtable{\par}{\if@noskipsec\mbox{}\fi\par}{}{}
\makeatother
\IfFileExists{footnotehyper.sty}{\usepackage{footnotehyper}}{\usepackage{footnote}}
\makesavenoteenv{longtable}
\usepackage{graphicx}
\makeatletter
\def\maxwidth{\ifdim\Gin@nat@width>\linewidth\linewidth\else\Gin@nat@width\fi}
\def\maxheight{\ifdim\Gin@nat@height>\textheight\textheight\else\Gin@nat@height\fi}
\makeatother
\setkeys{Gin}{width=\maxwidth,height=\maxheight,keepaspectratio}
\makeatletter
\def\fps@figure{htbp}
\makeatother
\ifLuaTeX
\usepackage[bidi=basic]{babel}
\else
\usepackage[bidi=default]{babel}
\fi
\babelprovide[main,import]{american}

\def\languageshorthands#1{}
\usepackage{hyperref}
\definecolor{linkcolor}{HTML}{D55E00}
\definecolor{citecolor}{HTML}{009E73}
\definecolor{urlcolor}{HTML}{0072B2}
\hypersetup{
    colorlinks,
    linkcolor={linkcolor},
    citecolor={citecolor},
    urlcolor={urlcolor}
}

\makeatletter
\@ifundefined{DecValTok}{}{
  \renewcommand{\DecValTok}[1]{\textcolor[HTML]{009E73}{#1}}
  \renewcommand{\FloatTok}[1]{\textcolor[HTML]{009E73}{#1}}
  \renewcommand{\ConstantTok}[1]{\textcolor[HTML]{009E73}{#1}}
  \renewcommand{\ControlFlowTok}[1]{\textcolor[HTML]{0072B2}{\textbf{#1}}}
  \renewcommand{\OtherTok}[1]{\textcolor[HTML]{000000}{#1}}
  \renewcommand{\CommentTok}[1]{\textcolor[HTML]{999999}{\textit{#1}}}
  \renewcommand{\AttributeTok}[1]{\textcolor[HTML]{CC79A7}{#1}}
  \renewcommand{\FunctionTok}[1]{\textcolor[HTML]{56B4E9}{#1}}
}
\makeatother

\usepackage{caption}
\usepackage{enumitem}
\setlist[itemize]{topsep=0pt}

\usepackage[vlined]{algorithm2e}
\usepackage{algorithmic}
\newcommand{\SD}{\operatorname{SD}}
\newcommand{\MAD}{\operatorname{MAD}}
\newcommand{\QAD}{\operatorname{QAD}}
\newcommand{\SQAD}{\operatorname{SQAD}}
\newcommand{\OQAD}{\operatorname{OQAD}}

\newcommand{\THDME}{\operatorname{THDME}}
\newcommand{\STHDME}{\operatorname{STHDME}}
\newcommand{\OTHDME}{\operatorname{OTHDME}}

\newcommand{\Q}{\operatorname{Q}}
\newcommand{\SM}{\operatorname{SM}}
\newcommand{\median}{\operatorname{median}}
\newcommand{\mean}{\operatorname{mean}}

\newcommand{\Pois}{\operatorname{Pois}}
\newcommand{\Exp}{\operatorname{Exp}}
\newcommand{\Pareto}{\operatorname{Pareto}(1, 1)}

\newcommand{\E}{\mathbb{E}}
\newcommand{\V}{\mathbb{V}}
\newcommand{\PR}{\mathbb{P}}
\usepackage{amsmath}
\usepackage{float}
\usepackage{csquotes}
\usepackage{booktabs}
\usepackage{longtable}
\usepackage{array}
\usepackage{multirow}
\usepackage{wrapfig}
\usepackage{colortbl}
\usepackage{pdflscape}
\usepackage{tabu}
\usepackage{threeparttable}
\usepackage{threeparttablex}
\usepackage[normalem]{ulem}
\usepackage{makecell}
\usepackage{xcolor}
\ifLuaTeX
  \usepackage{selnolig}  
\fi
\usepackage[style=alphabetic,sorting=anyt]{biblatex}
\IfFileExists{bookmark.sty}{\usepackage{bookmark}}{\usepackage{hyperref}}
\IfFileExists{xurl.sty}{\usepackage{xurl}}{} 
\hypersetup{
  pdftitle={Quantile absolute deviation},
  pdfauthor={Andrey Akinshin},
  pdflang={en-US},
  colorlinks=true,
  linkcolor={linkcolor},
  filecolor={Maroon},
  citecolor={citecolor},
  urlcolor={urlcolor},
  pdfcreator={LaTeX via pandoc}}

\title{Quantile absolute deviation}
\author{Andrey Akinshin\\
Huawei Research, \href{mailto:andrey.akinshin@gmail.com}{\nolinkurl{andrey.akinshin@gmail.com}}}
\date{}

\begin{document}
\maketitle
\begin{abstract}
The median absolute deviation (MAD) is a popular robust measure of statistical dispersion.
However, when it is applied to non-parametric distributions (especially multimodal, discrete, or heavy-tailed),
lots of statistical inference issues arise.
Even when it is applied to distributions with slight deviations from normality and these issues are not actual,
the Gaussian efficiency of the MAD is only 37\% which is not always enough.

In this paper, we introduce the \emph{quantile absolute deviation} (QAD) as a generalization of the MAD.
This measure of dispersion provides a flexible approach to analyzing properties of non-parametric distributions.
It also allows controlling the trade-off between robustness and statistical efficiency.
We use the trimmed Harrell-Davis median estimator based on the highest density interval of the given width
as a complimentary median estimator that gives
increased finite-sample Gaussian efficiency compared to the sample median
and a breakdown point matched to the QAD.

As a rule of thumb, we suggest using two new measures of dispersion
called the \emph{standard QAD} and the \emph{optimal QAD}.
They give 54\% and 65\% of Gaussian efficiency having breakdown points of 32\% and 14\% respectively.

\end{abstract}


\hypertarget{introduction}{%
\section{Introduction}\label{introduction}}

The \emph{median absolute deviation around the median} (\(\MAD\)) is a widely used measure of statistical dispersion
which is often used as a robust alternative to the standard deviation.
For a sample \(X = \{ X_1, X_2, \ldots, X_n \}\) of i.i.d. random variables, it is given by

\[
\MAD(X) = C \cdot \median(|X - \median(X)|),
\]

where \(C\) is a scale constant that can be used to make the \(\MAD\) Fisher-consistent (\autocite{fisher1922})
for the standard deviation under the normal distribution.
One of the first notes about the \(\MAD\) can be found in \autocite[p388]{hampel1974} where it is attributed to Gauss:
``It was, in fact, mentioned briefly by Gauss (\autocite{gauss1816}), who noted its simplicity,
but with the same breath, dismissed it because of
its low asymptotic efficiency of about 40 percent for strictly normal data (cf.~also \autocite{stigler1973a}).''
Despite the remark about poor statistical efficiency, the \(\MAD\) has become a popular robust scale estimator
due to its high breakdown point of \(50\%\).
It is discussed as a robust measure of dispersion in various statistical textbooks including
\autocite{wilcox2016,huber2009,maronna2019,mosteller1977,hampel1986,jureckova2019}.

The actual asymptotic Gaussian efficiency of the \(\MAD\) is only \(36.75\%\).
Using adjusted values of \(C\), we can make the \(\MAD\) consistent with the standard deviation for finite samples.
The corresponding bias-correction factors can be found in \autocite{park2020}.
Its finite-sample Gaussian efficiency is a little bit higher than its asymptotic value,
but the difference is not dramatic (when the sample size is larger than \(14\), the efficiency is below \(40\%\)).
There are various ways to improve efficiency.
For example, we can consider an alternative median estimator.
Conventionally, the \(\MAD\) is based on the sample median
(when \(n\) is odd, the median is the middle order statistic;
when \(n\) is even, the median is the arithmetic average of the two middle order statistics)
which is not the most robust median estimator.
It can be replaced by the Harrell-Davis median estimator (see \autocite{harrell1982})
which is asymptotically consistent with the sample median (see \autocite{yoshizawa1985}).
Strictly speaking, this estimator is not robust since it is based on a weighted sum of all order statistics.
However, small and large order statistics get negligible weight coefficients
which make this approach practically acceptable in the case of non-extreme outliers.
In order to increase robustness,
we can also consider the trimmed modification of the Harrell-Davis median estimator
which has a higher breakdown point and slightly lower efficiency (see \autocite{akinshin2022thdqe}).
The finite-sample bias correction factors for both approaches are given in \autocite{akinshin2022madfactors}.

There is also a plethora of various alternatives to the \(\MAD\) like
the interquartile and interdecile ranges,
the trimmed standard deviation (see \autocite{lax1985}),
the Winsorized standard deviation (see \autocite{wilcox2016}),
the Rousseeuw and Croux \(S_n\) and \(Q_n\) scale estimators (see \autocite{rousseeuw1993}),
the biweight midvariance (see \autocite{lax1985}),
the Shamos estimator (see \autocite{shamos1976}),
and others (e.g., see \autocite{daniell1920}).
Each estimator has its own properties including breakdown point, statistical efficiency,
componential efficiency, influence function (see \autocite{hampel1974}), etc.

In this paper, we build a generalization of the \(\MAD\)
that we call the \emph{quantile absolute deviation around the median} (\(\QAD\)).
It is defined as follows:

\[
\QAD(X, p) = K \cdot \Q(|X - \median(X)|, p),
\]

where \(\Q(\cdot, p)\) is a quantile estimator,
\(K\) is a scale constant,
\(p \in [0; 1]\).
The parameter \(p\) allows customizing the trade-off between statistical efficiency and robustness.
The \(\MAD\) is a special case of the \(\QAD\): \(\MAD(X) = \QAD(X, 0.5)\).
When we increase \(p\) from \(0.5\) to \(1.0\), we trade robustness (the breakdown point becomes \(1-p\))
for increased efficiency.

The \(\MAD\) is often used together with the \(\median\)
since the interval from \(\median(X)-\MAD(X)\) to \(\median(X)+\MAD(X)\) covers \(50\%\) of the distribution.
When we use \(\QAD\), the interval from \(\median(X)-\QAD(X, p)\) to \(\median(X)+\QAD(X, p)\)
covers \(p \cdot 100\%\) of the distribution.
Thus, we have flexibility in describing distribution properties.
Since the breakdown point of \(\QAD(X, p)\) is \(1-p\),
we can use an alternative median estimator with the same breakdown point
to obtain better efficiency for the median estimations.
We suggest using the trimmed Harrell-Davis median estimator
based on the highest density interval of size \(p\) (see \autocite{akinshin2022thdqe}).

In practice, the breakdown point of \(50\%\) is not always required.
For example, in \autocite[p.26--28]{hampel1986},
it is stated that we can expect about \(1\text{--}10\%\) of gross errors in real data sets.
Therefore, based on the knowledge of the data source, one can use high values of \(p\) up to \(0.8\text{--}0.9\).
In this paper, we discuss two rules of thumb for choosing the value of \(p\).
The first one suggests using \(p=\Phi(1)-\Phi(-1)\approx 68.27\%\) (the corresponding Gaussian efficiency is \(54.06\%\)).
The intuition behind this value is that the corresponding interval
\([\median(X)-\QAD(X, p); \median(X)+\QAD(X, p)]\) converges to \([\mu-\sigma; \mu+\sigma]\) under normality.
The second one is \(p \approx 86.17\%\)
since it is the point at which \(\QAD\) achieves its highest Gaussian efficiency of \(65.22\%\).
We denote them by the standard \(\QAD\) (\(\SQAD\)) and the optimal \(\QAD\) (\(\OQAD\)).
The corresponding median estimators are also discussed.

While the \(\MAD\), the \(\SQAD\), and the \(\OQAD\) are reliable and robust measures of statistical dispersion
for unimodal continuous light-tailed distributions with slight deviations from normality,
they can be misleading in the non-parametric case.
Discretization may lead to zero dispersion values,
multimodality may lead to unstable estimations,
heavy-tailedness may lead to violated assumptions typical for the normal model.
In order to get a broad perspective of dispersion properties of the underlying non-parametric distribution,
we can consider the \(\QAD(X, p)\) function for all \(p \in [0; 1]\).

In this paper, we discuss various aspects of using the \(\QAD\) and its complimentary median estimator:
robustness, statistical efficiency, unbiasedness, and possible issues with non-parametric distributions.

The paper is organized as follows.
In Section~\ref{sec:mad},
we revise the basic properties of the \(\MAD\)
and discuss caveats of using the \(\MAD\) with non-parametric distributions.
In Section~\ref{sec:qad},
we discuss the concept of the \(\QAD\),
calculate the asymptotic consistency constants and Gaussian efficiency,
and introduce the \(\SQAD\) and the \(\OQAD\).
In Section~\ref{sec:thdme},
we consider a complimentary median estimator for \(\QAD\) which has the same breakdown point
and discuss its special cases for the \(\SQAD\) and the \(\OQAD\).
In Section~\ref{sec:sim}, we perform a series of numerical simulations
to get the values of the finite-sample consistency constant values
and Gaussian efficiency values of the presented estimators.
In Section~\ref{sec:summary}, we summarize all the results.
In Appendix~\ref{sec:refimpl}, we provide a reference R implementation of the discussed approach.

\clearpage

\hypertarget{sec:mad}{%
\section{Median absolute deviation}\label{sec:mad}}

For a sample \(X = \{ X_1, X_2, \ldots, X_n \}\) of i.i.d. random variables,
the \emph{median absolute deviation} is given by

\[
\MAD(X) = \median(|X - \median(X)|).
\]

In order to make it an unbiased Fisher-consistent estimator for the standard deviation (\(\SD\)),
we should introduce a consistency constant \(C_n\):

\[
\MAD_n(X) = C_n \cdot \median(|X - \median(X)|).
\]

The asymptotic value of \(C_n\) is well-known: \(C_{\infty} = 1 / \Phi^{-1}(0.75) \approx 1.4826\)
where \(\Phi^{-1}\) is the quantile function of the standard normal distribution
(a detailed proof can be found in \autocite[Section 2.1]{akinshin2022madfactors}).
The finite-sample bias-correction factors can be found in \autocite{park2020}.

One of the most popular measures of estimator robustness is the breakdown point.
It describes the portion of the distribution that can be replaced by arbitrary large values
without corrupting the obtained estimations.
The \(\MAD\) is highly robust, its breakdown point is \(50\%\)
which is the highest possible value for a scale estimator (see \autocite[p14]{rousseeuw1987}).

\begin{figure}[ht!]

{\centering \includegraphics{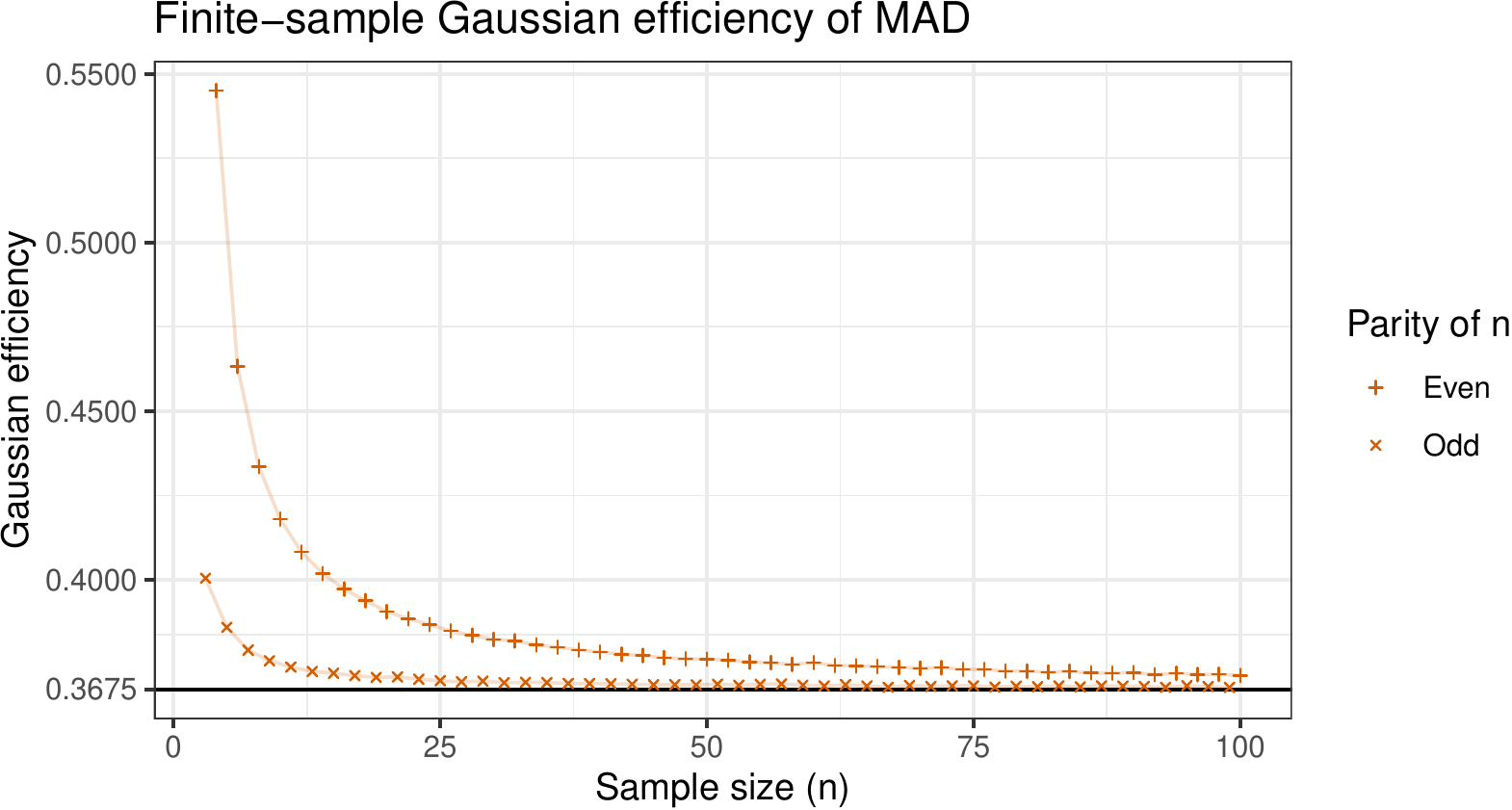} 

}

\caption{Finite-sample Gaussian efficiency of the median absolute deviation.}\label{fig:mad-efficiency}
\end{figure}

Its asymptotic relative statistical efficiency to the standard deviation under normality (\emph{Gaussian efficiency})
is \(36.75\%\) (the proof is in Section~\ref{sec:qad-age}).
However, for finite samples, the statistical efficiency is higher as shown in Figure~\ref{fig:mad-efficiency}.
The raw values are presented in Table~\ref{tab:tab-scale-efficiency}.

When researchers use the \(\MAD\) as a robust replacement for the standard deviation,
they typically use assumptions that are valid only for the normal distribution
and some unimodal continuous light-tailed distributions with slight deviations from normality.
Unfortunately, these assumptions can be violated
when the considered distribution is multimodal, discrete, heavy-tailed, or has high deviations from normality.
In this section, we discuss the caveats of using the \(\MAD\) with non-parametric distributions.

\clearpage

\hypertarget{sec:mad_multimodal}{%
\subsection{MAD and multimodal distributions}\label{sec:mad_multimodal}}

When non-robust estimators are used, a single corrupted element can easily distort the estimation.
A transition towards robust approaches brings a lot of benefits in terms of stability:
a single altered element cannot introduce an extreme change in the estimation value.
Having this knowledge, many researchers typically expect that robust estimations would fit a narrow range of values
from an approximately normal sampling distribution.
Therefore, they can omit the phase of exploring the distribution of estimations
and draw conclusions based on a single trial assuming
that all possible estimations are close enough to each other.
However, while robust estimators indeed provide a decent defense against gross errors,
they still can have a wide range of possible values.
The corresponding sampling distribution can also have heavy deviations from normality.

\begin{figure}[ht!]

{\centering \includegraphics{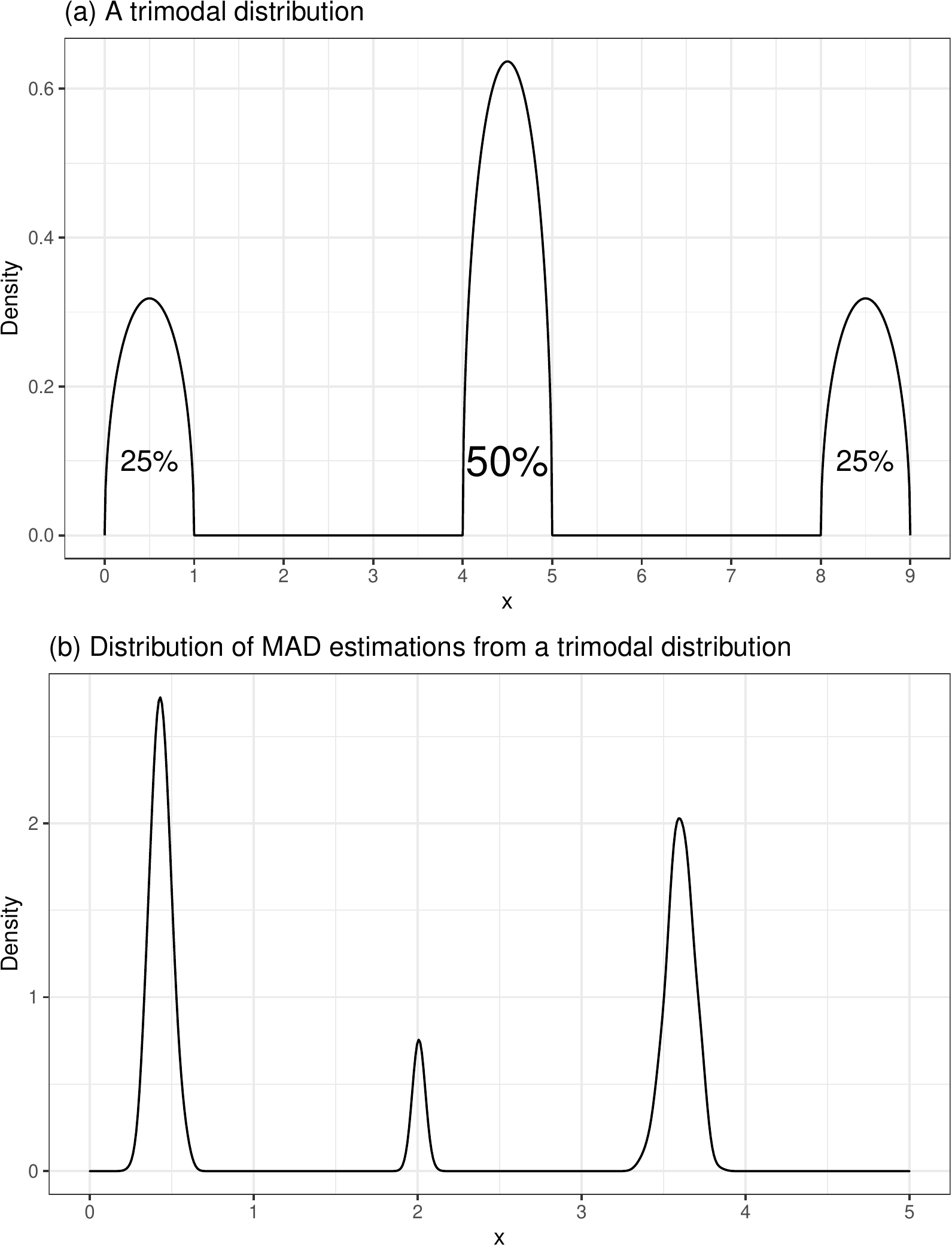} 

}

\caption{A trimodal distribution and the corresponding sampling distribution of the MAD.}\label{fig:mad-multimodal}
\end{figure}

One of the most severe problems arises in the multimodal case.
Let us consider a trimodal distribution that is shown in Figure~\ref{fig:mad-multimodal}a.
This distribution has three non-intersecting intervals as shown in Table~\ref{tab:mad-trimodal}.

\begin{longtable}[]{@{}rr@{}}
\caption{\label{tab:mad-trimodal} Three modes of the distribution from Figure~\ref{fig:mad-multimodal}a.}\tabularnewline
\toprule()
Interval & Portion \\
\midrule()
\endfirsthead
\toprule()
Interval & Portion \\
\midrule()
\endhead
\([0;1]\) & \(25\%\) \\
\([4;5]\) & \(50\%\) \\
\([8;9]\) & \(25\%\) \\
\bottomrule()
\end{longtable}

While the distribution has an unambiguously defined median \(M = 4.5\),
its quantile function of the absolute deviations around the median (\(|X - \median(X)|\))
has a discontinuity at 0.5.
Therefore, we cannot unambiguously define the \(\MAD\) value.
Indeed, the \([M-\MAD; M+\MAD]\) interval should cover exactly \(50\%\) of the distribution.
In the considered case, there are multiple ways to define such an interval.
The narrowest and the widest suitable intervals are \([4;5]\) to \([1;8]\) respectively.

Now we explore the practical implications of working with such a distribution.
Let us take \(1\,000\) random samples of size \(100\) from this distribution,
estimate the \(\MAD\) value for each sample,
and build a new distribution based on the obtained estimations.
The density plot of the observed sampling distribution is presented in Figure~\ref{fig:mad-multimodal}b
(we use the kernel density estimation with the normal kernel and
the Sheather \& Jones method to select the bandwidth, see \autocite{sheather1991}).

As we can see, the sampling distribution of the \(\MAD\) is also trimodal.
In the general case, the distance between these modes has the same magnitude as
the gap around the \(0.5^\textrm{th}\) quantile of \(|X - \median(X)|\).

This problem can be viewed from another perspective.
For a distribution with probability density function \(f\) and the true median value \(M\),
the distribution of the sample median estimations is asymptotically normal
(according to \autocite{stigler1973b} it was firstly derived by Laplace)
with mean \(M\) and variance

\[
\frac{1}{4nf(M)^2}.
\]

As we can see, the variance equation requires \(f(M)>0\) which is not always true in the general case.
Therefore, if we have discontinuities around the median of \(X\) or \(|X - \median(X)|\),
the sampling distribution of the \(\MAD\) may have a multimodal form with a wide range of values.

Even if \(f(M)>0\), the sampling distribution of the median is only \emph{asymptotically} normal,
which means that we can expect noticeable deviations from normality when the sample size is small (see \autocite{rider1960}).

Thus, despite the high robustness of the \(\MAD\),
the obtained estimations can significantly vary between samples
and we cannot speculate on the form of the corresponding sampling distributions based on a few samples.

\clearpage

\clearpage

\hypertarget{sec:mad_discrete}{%
\subsection{MAD and discrete distributions}\label{sec:mad_discrete}}

When we work with continuous distributions, we typically assume that the dispersion value is strictly positive.
Thanks to this property, dispersion measures are often used as denominators in various statistical equations
(e.g., in the standard score, in effect size measures like the Cohen's d,
in null hypothesis significance tests like the Student's t-test, and so forth).
In theory, the probability of obtaining tied values from a continuous distribution is zero
so that we should not expect division by zero.
In practice, tied values can arise in the continuous case due to the limited resolution of the measurement device.
Of course, we can also expect tied values in samples from
discrete distributions (e.g., Bernoulli distribution or Binomial distribution)
or mixtures of discrete and continuous distributions (e.g., the rectified Gaussian distribution, see \autocite{socci1997}).

\begin{figure}[ht!]

{\centering \includegraphics{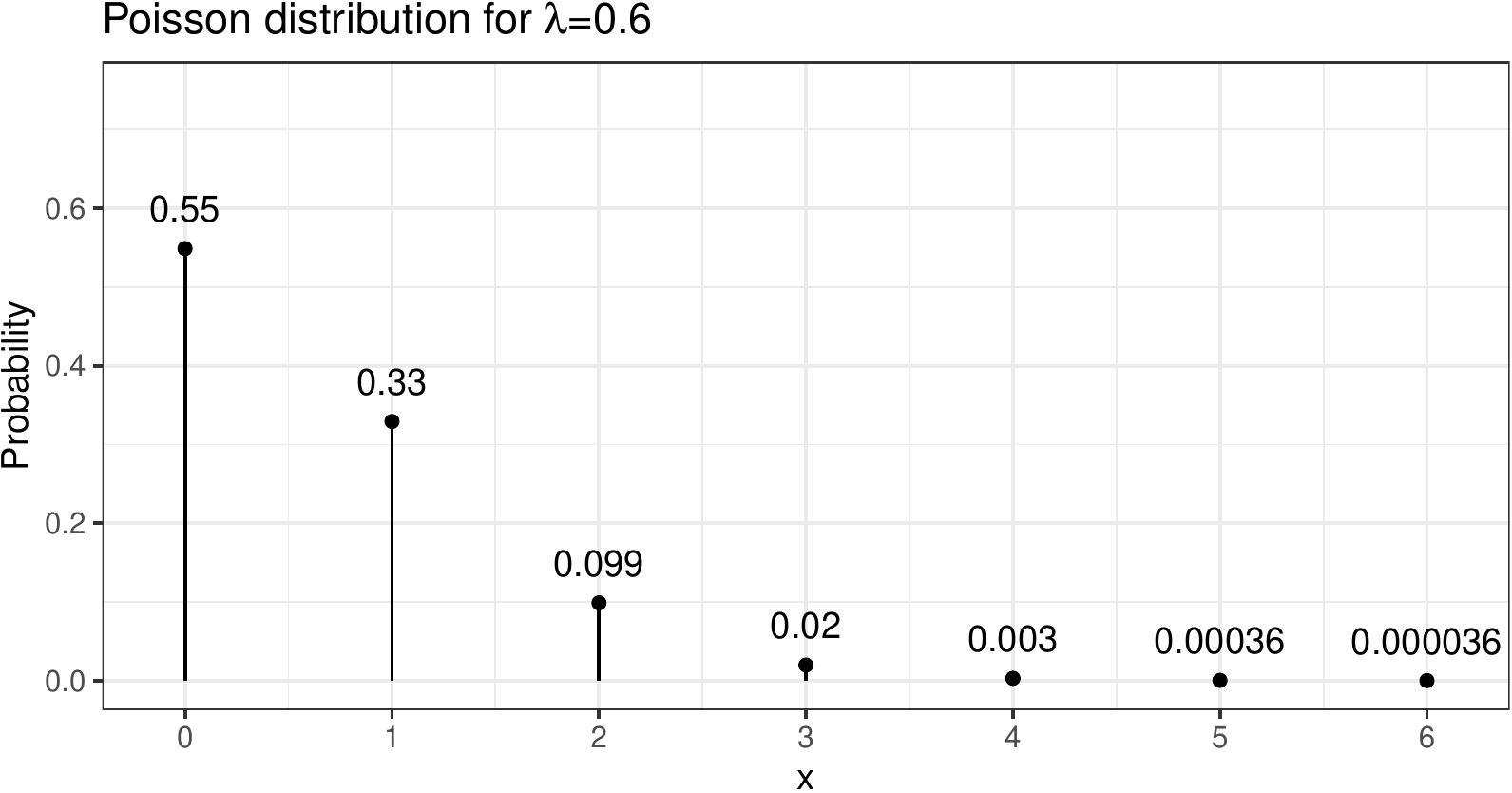} 

}

\caption{Probability mass function of the Poisson distribution for $\lambda=0.6$.}\label{fig:mad-discrete}
\end{figure}

As an example of a discrete distribution, let us consider the Poisson distribution \(\Pois(\lambda)\).
Its probability mass function is defined by \(p(k)=\lambda^k e^{-\lambda} / k!\).
It is easy to see that when \(\lambda < \lambda_0 = -\ln(0.5) \approx 0.6931\), we have \(p(0) > 0.5\)
(e.g., in Figure~\ref{fig:mad-discrete}, \(\Pois(0.6)\) is presented; \(p(0) \approx 0.55\)).
In this case, more than half of the distribution elements are equal to zero, which gives a zero value of the \(\MAD\).
Such a situation can become a severe issue in statistical inference.

Similarly to the breakdown point, we can introduce the \emph{degenerate point} of a dispersion estimator that describes
the minimum portion of the sample that should be replaced by zeros to get zero estimation.
For example,
the asymptotic degenerate points of the \(\MAD\), the interquartile range,
the interdecile range, and the standard deviation are
\(50\%\), \(50\%\), \(80\%\), and \(100\%\) respectively.

In practice, the dispersion estimator can be redefined by introducing an artificial lower bound
so that it never becomes zero.
For example, we can consider a modification of \(\MAD\) defined by \(\MAD^*(X) = \max(\MAD(X), \MAD_{\min})\),
where \(\MAD_{\min} > 0\).
In this case, \(\MAD_{\min}\) becomes the new ``degenerate'' value.
We can correspondingly redefine the degenerate point so that it describes
the minimum portion of the sample that should be replaced by zeros to get the minimum possible estimation value.

The degenerate point plays an important role when we choose a proper dispersion estimator for discrete distributions
or mixtures of discrete and continuous distributions
since it defines the domain in which considered estimators can be actually used.
For example, if we work with the Poisson distribution \(\Pois(\lambda)\), we need a dispersion estimator
with a degenerate point greater than \(p(0) = e^{-\lambda}\).

\clearpage

\hypertarget{sec:mad_heavy}{%
\subsection{MAD and heavy-tailed distributions}\label{sec:mad_heavy}}

When we work the normal distribution \(\mathcal{N}(\mu, \sigma^2)\), its structure can be described only by two parameters:
the mean \(\mu\) and the standard deviation \(\sigma\).
This representation is powerful
because we can instantly get various insights about the distribution based solely on these two values.
A widely used example of assumptions related to the normal model is the 68--95--99.7 rule.
It says that intervals \([\mu-\sigma;\mu+\sigma]\), \([\mu-2\sigma;\mu+2\sigma]\), and \([\mu-3\sigma;\mu+3\sigma]\)
cover \(68\%\), \(95\%\), and \(99.7\%\) of the normal distribution respectively.
This rule is linked with the three-sigma rule of thumb that implies that the interval \([\mu-3\sigma;\mu+3\sigma]\)
covers almost all the distribution values.
While this empirical rule is applicable to numerous unimodal continuous light-tailed distributions,
it can be easily violated in the case of heavy-tailed distributions.

\begin{figure}[ht!]

{\centering \includegraphics{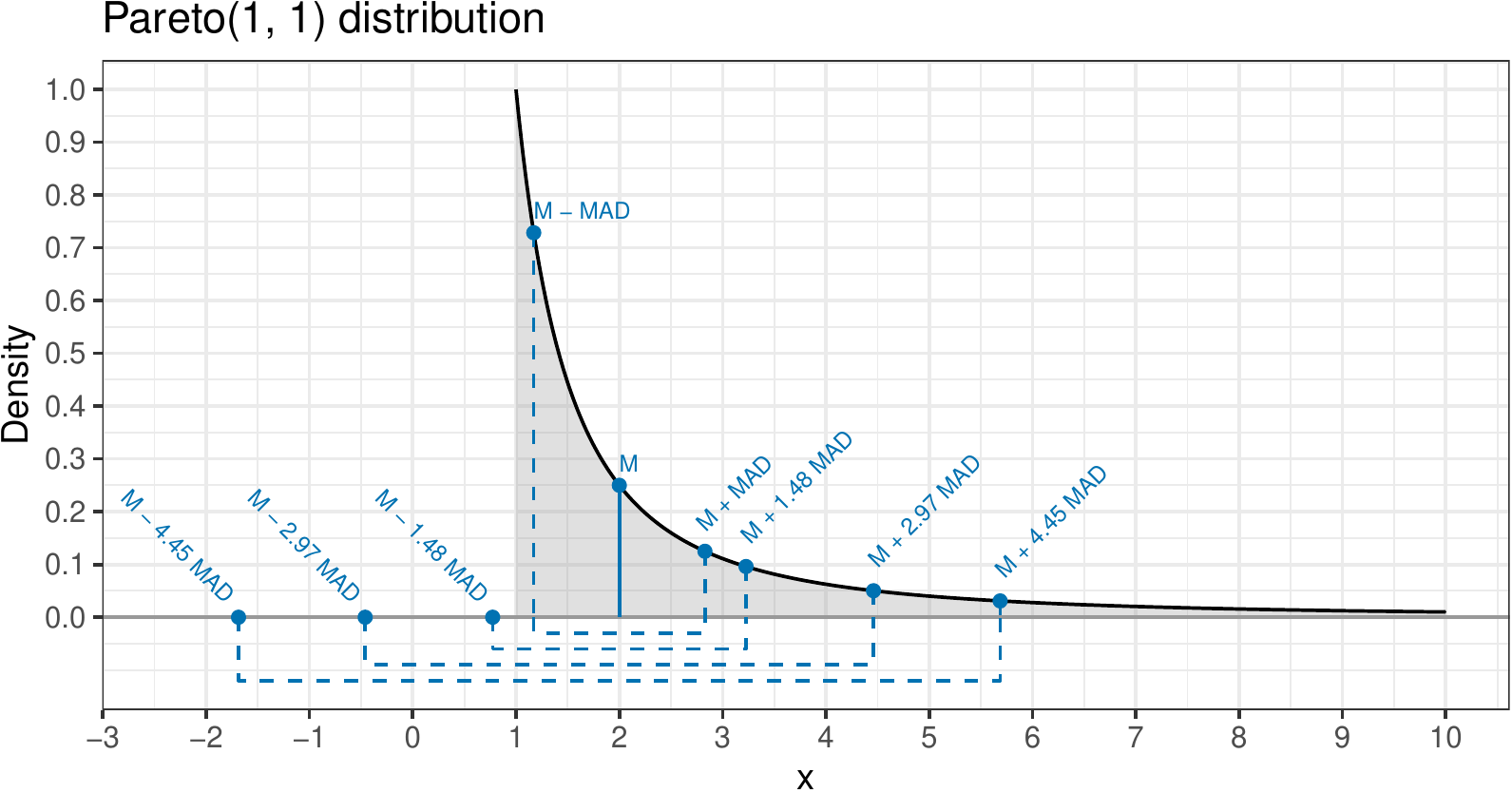} 

}

\caption{The density plot of the Pareto(1, 1) distribution.}\label{fig:mad-heavy}
\end{figure}

Let us consider the \(\Pareto\) distribution which is a commonly used example of a heavy-tailed distribution.
Its true median value is \(M = 2\) and the true median absolute deviation value is \(\MAD \approx 0.83\).
The corresponding density plot is presented in Figure~\ref{fig:mad-heavy}.

The variance of this distribution is infinite, therefore we cannot use the standard deviation as a measure of dispersion.
However, let us check what would happen with the 68--95--99.7 rule if we try to apply it
using a scaled \(\MAD\) as a standard deviation estimator.
We consider intervals \([M-k \cdot \MAD; M+k \cdot \MAD]\) for
\(k \in \{ 1, C_{\infty}, 2C_{\infty}, 3C_{\infty} \}\).
The actual coverage values of all intervals are presented in Table~\ref{tab:mad-coverage}.

\begin{table}[!h]

\caption{\label{tab:mad-coverage}Coverage of the Normal(0, 1) and Pareto(1, 1) distributions by various intervals.}
\centering
\begin{tabular}[t]{c|>{\centering\arraybackslash}p{4cm}|>{\centering\arraybackslash}p{4cm}}
\hline
\multicolumn{1}{c|}{k} & \multicolumn{2}{c}{$\mathbb{P}(M - k\cdot \mathrm{MAD}) \leq X \leq \mathbb{P}(M + k\cdot \mathrm{MAD})$} \\
\cline{1-1} \cline{2-3}
 & Normal & Pareto(1, 1)\\
\hline
1.00 & 0.500 & 0.500\\
\hline
1.48 & 0.682 & 0.690\\
\hline
2.97 & 0.955 & 0.776\\
\hline
4.45 & 0.997 & 0.824\\
\hline
\end{tabular}
\end{table}

As we can see, a blind usage of the three-sigma rule in the heavy-tailed case can lead to misleading insights.
In the above example, the interval \([M- 3C_{\infty} \cdot \MAD; M+ 3C_{\infty} \cdot \MAD]\)
actually cover only \(82.4\%\) of the \(\Pareto\) distribution instead of the typical \(99.7\%\) for the normal one.

\clearpage

\hypertarget{sec:qad}{%
\section{Quantile absolute deviation}\label{sec:qad}}

Now we introduce a generalization of the \(\MAD\)
that we call the \emph{quantile absolute deviation around the median} (\(\QAD\)).
We define the \(\QAD\) by

\[
\QAD(X, p) = \Q(|X - \median(X)|, p),
\]

where \(Q(\cdot, p)\) is a quantile estimator, \(p\) is the order of the target quantile.
In the scope of this paper,
we are using the Hyndman-Fan Type 7 quantile estimator (see \autocite{hyndman1996})
which is the most popular traditional quantile estimator based on one or two order statistics
(it is used by default in R, Julia, NumPy, and Excel).
It is given by

\[
Q(X, p) = X_{(\lfloor h \rfloor)}+(h-\lfloor h \rfloor)(X_{(\lceil h \rceil)}-X_{(\lfloor h \rfloor)}),
\quad h = (n-1)p+1,
\]

where \(\lfloor \cdot \rfloor\) and \(\lceil \cdot \rceil\) are the floor and ceiling functions,
\(X_{(i)}\) is the \(i^\textrm{th}\) order statistic of \(X\).
It is easy to see that such an estimator is consistent with the sample median: \(\Q(X, 0.5) = \median(X)\).
Therefore, the \(\MAD\) is a special case of \(\QAD\):

\[
\MAD(X) = \median(|X-\median(X)|) = \Q(|X - \median(X, 0.5)|, 0.5) = \QAD(X, 0.5).
\]

While the interval

\[
[\median(X) - \MAD(X);\; \median(X) + \MAD(X)]
\]

covers \(50\%\) of the distribution, the interval

\[
[\median(X) - \QAD(X,p);\; \median(X) + \QAD(X,p)]
\]

covers \(p\cdot 100\%\) of the distribution.
This property gives us a broader view of the distribution dispersion
via exploring the \(\QAD(X, p)\) function for all \(p \in [0; 1]\).
Therefore, we can pick such a value of \(p\) that helps to avoid \(\MAD\)-specific problems listed in the previous section.
The parameter \(p\) affects the statistical efficiency, the breakdown point (which equals \(1-p\)),
the degenerate point (which equals \(p\)), the influence function, and other properties of \(\QAD(X, p)\).

Now we will
explore examples of the \(\QAD\) for different distributions (Section~\ref{sec:qad-ex}),
derive the asymptotic values of consistency constants (Section~\ref{sec:qad-cc})
and Gaussian efficiency (Section~\ref{sec:qad-age}),
and introduce the standard \(\QAD\) (Section~\ref{sec:qad-sqad})
and the optimal \(\QAD\) (Section~\ref{sec:qad-oqad}).

\clearpage

\hypertarget{sec:qad-ex}{%
\subsection{Examples of the quantile absolute deviation functions}\label{sec:qad-ex}}

\begin{figure}[ht!]

{\centering \includegraphics{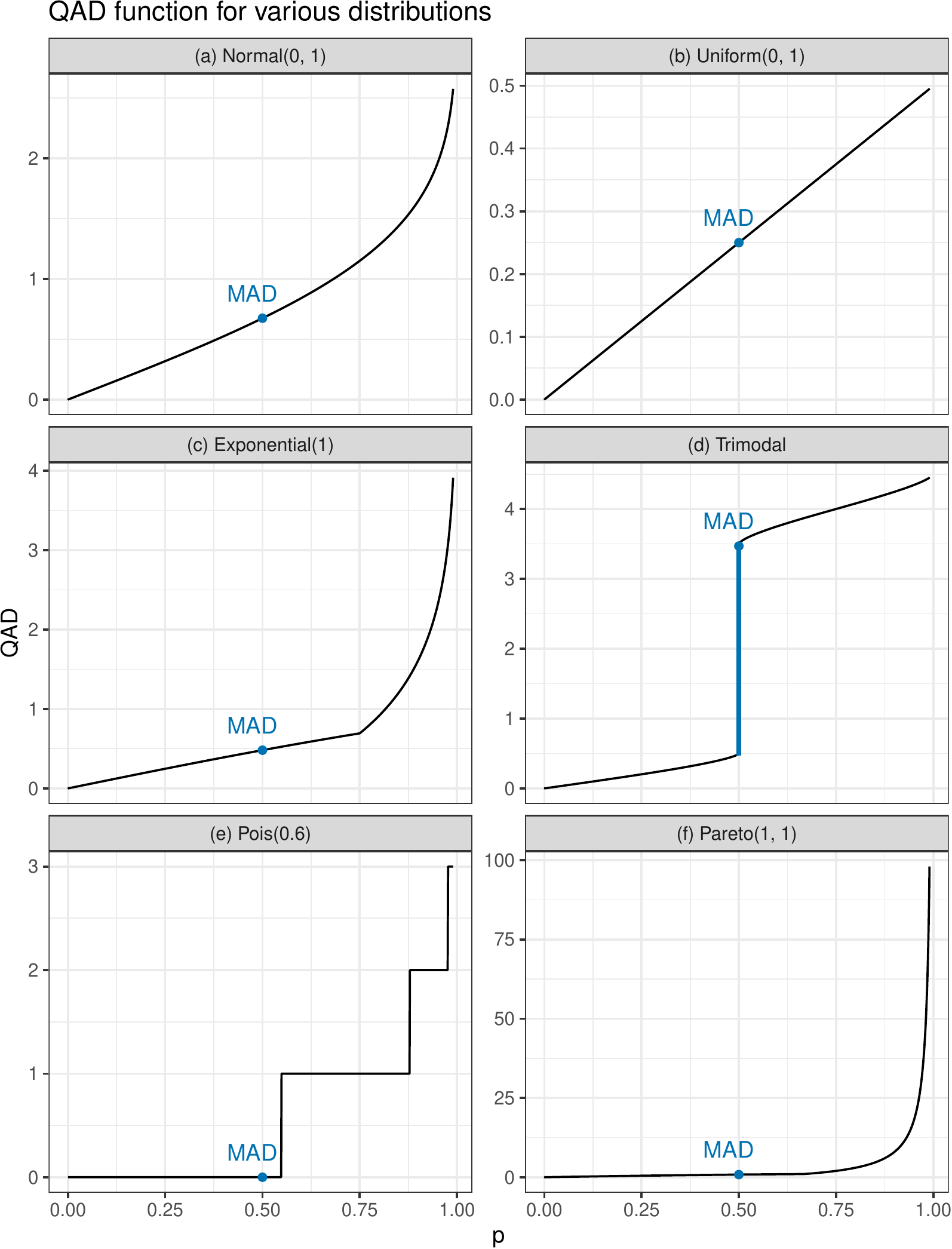} 

}

\caption{QAD function for various distributions.}\label{fig:qad-functions}
\end{figure}

In order to get a better understanding of the \(\QAD(X, p)\) function behavior,
we consider several examples of this function (Figure~\ref{fig:qad-functions}) for various distributions.

For some of these distributions,
we derive the exact asymptotic equation for its quantile absolute deviation as follows.
Let us denote the distribution cumulative distribution function (CDF) by \(F\) and the true distribution median by \(M\).
For simplification, we denote the asymptotic value of \(\QAD(X, p)\) by \(v_p\):

\[
v_p = \lim_{n \to \infty} \E[\Q(|X-M|, p)].
\]

By the definition of quantiles, this can be rewritten as:

\[
\PR(|X_1 - M| < v_p) = p,
\]

which is the same as

\[
\PR(-v_p < X_1 - M < v_p) = p.
\]

Hence,

\[
\PR(M - v_p < X_1 < M + v_p) = p.
\]

This can be rewritten using the CDF:

\begin{equation}
F(M + v_p) - F(M - v_p) = p.
\label{eq:qad-solve}
\end{equation}

The solution of this equation gives us the asymptotic expected value of \(\QAD(X, p)\).

We also consider the situation in which we have the minimum \(x\) value \(x_{\min}\) so that \(F(x) = 0\) for \(x \leq x_{\min}\).
In this case, it is necessary to consider two cases: \(M - v_p \leq x_{\min}\) and \(M - v_p > x_{\min}\).
The critical value \(v_p^*\) is defined by \(M-v_p^* = x_{\min}\), which gives us \(v_p^* = M - x_{\min}\).
The corresponding critical value \(p^*\) can be derived from Equation~\eqref{eq:qad-solve}:

\begin{equation}
p^* = F(M + v_p^*) - F(M - v_p^*) = F(2M - x_{\min}) - F(x_{\min}) = F(2M - x_{\min}).
\label{eq:qad-pstar}
\end{equation}

Once the critical value \(p^*\) is obtained, we should consider cases \(p \leq p^*\) and \(p > p^*\) independently.
For the case \(p \leq p^*\), Equation~\eqref{eq:qad-solve} should be used.
For the case \(p > p^*\), we can omit \(F(M - v_p)\) since it is always zero so that we should solve

\begin{equation}
F(M + v_p) = p.
\label{eq:qad-solve2}
\end{equation}

\clearpage

\hypertarget{standard-normal-distribution}{%
\subsubsection{Standard normal distribution}\label{standard-normal-distribution}}

We consider the standard normal distribution \(\mathcal{N}(0, 1)\)
given by \(F(x)=\Phi(x)\) with the median value \(M=0\).
Equation~\eqref{eq:qad-solve} has the following form:

\[
\Phi(v_p) - \Phi(-v_p) = p.
\]

Using \(\Phi(-v_p) = 1 - \Phi(v_p)\), we get:

\[
\Phi(v_p) = \frac{p+1}{2},
\]

which is the same as

\[
v_p = \Phi^{-1} \Big( \frac{p+1}{2} \Big).
\]

Thus, if \(X \sim \mathcal{N}(0, 1)\),

\begin{equation}
\lim_{n \to \infty} \E[\QAD(X, p)] = \Phi^{-1} \Big( \frac{p+1}{2} \Big).
\label{eq:qad-normal}
\end{equation}

The corresponding plot is presented in Figure~\ref{fig:qad-functions}a.

We can use this \(\QAD\) function as a reference mental model
in which all the assumptions related to the normal distribution are satisfied.

\hypertarget{standard-uniform-distribution}{%
\subsubsection{Standard uniform distribution}\label{standard-uniform-distribution}}

We consider the standard uniform distribution \(\mathcal{U}(0, 1)\)
given by \(F(x)=x\) on \([0;1]\) with the median value \(M=0.5\).
Equation~\eqref{eq:qad-solve} has the following form:

\[
(0.5 + v_p) - (0.5 - v_p) = p,
\]

which is the same as

\[
v_p = p / 2.
\]

Thus, if \(X \sim \mathcal{U}(0, 1)\),

\[
\lim_{n \to \infty} \E[\QAD(X, p)] = \frac{p}{2}.
\]

The corresponding plot is presented in Figure~\ref{fig:qad-functions}b.

Since \(\mathcal{U}(0, 1)\) is a continuous unimodal light-tailed distribution,
the plot of the asymptotic \(\QAD(X, p)\) values is quite similar to the normal one
because we have only slight deviations from normality.
Most of the assumptions related to the normal distributions are approximately satisfied.

\clearpage

\hypertarget{standard-exponential-distribution}{%
\subsubsection{Standard exponential distribution}\label{standard-exponential-distribution}}

We consider the standard exponential distribution \(\Exp(1)\)
given by \(F(x)=1 - e^{-x}\) with the median value \(M=\ln 2\).
Since \(F\) is defined only for \(x \geq x_{\min} = 0\), we have to consider two cases: \(M - v_p \leq 0\) and \(M - v_p > 0\).
The critical value \(v_p^*\) is defined by \(v_p^* = M - x_{\min} = \ln 2\).
From Equation~\eqref{eq:qad-pstar}, we obtain \(p^*\):

\[
p^* = F(2M - x_{\min}) = F(2M) = 1 - e^{-2\ln 2} = 1 - 0.25 = 0.75.
\]

Let us consider the first case when \(p \leq p^* = 0.75\).
Equation~\eqref{eq:qad-solve} has the following form:

\[
(1 - e^{-\ln 2 - v_p}) - (1 + e^{-\ln 2 + v_p}) = p,
\]

which is the same as

\[
e^{v_p} - e^{-v_p} = 2p.
\]

By multiplying both sides of the equation by \(e^{v_p}\), we get:

\[
(e^{v_p})^2 - 2p \cdot (e^{v_p}) - 1 = 0.
\]

This is a quadratic equation for \(e^{v_p}\) with a solution given by

\[
e^{v_p} = \frac{2p \pm \sqrt{4p^2 + 4}}{2} = p \pm \sqrt{p^2+1}.
\]

Since \(e^{v_p}\) is always positive, only the plus is applicable for \(\pm\).
Taking the natural logarithm from both parts, we get the result:

\[
v_p = \ln(p + \sqrt{p^2+1}).
\]

Now let us consider the second case when \(p > p^* = 0.75\).
Equation~\eqref{eq:qad-solve2} has the following form:

\[
(1 - e^{-\ln 2 - v_p}) = p,
\]

which is the same as

\[
e^{-\ln 2 - v_p} = 1 - p.
\]

Taking the natural logarithm from both parts, we can express \(v_p\):

\[
v_p = -\ln 2 - \ln (1 - p).
\]

Thus, if \(X \sim \Exp(1)\),

\[
\lim_{n \to \infty} \E[\QAD(X, p)] = \begin{cases}
\ln(p + \sqrt{p^2+1}), & \textrm{if}\; p \leq 0.75,\\
-\ln 2 - \ln (1 - p), & \textrm{if}\; p > 0.75.
\end{cases}
\]

The corresponding plot is presented in Figure~\ref{fig:qad-functions}c.

This means that within the \(75\%\) interval around the median, we can use normal distribution assumptions.
If we are interested in distribution tails, we need non-parametric approaches.

\clearpage

\hypertarget{trimodal-distribution}{%
\subsubsection{Trimodal distribution}\label{trimodal-distribution}}

The trimodal distribution from Figure~\ref{fig:mad-multimodal}a can be described using the following CDF:

\[
F(x) \in \begin{cases}
\{ 0.00 \},   & \quad \textrm{if}\, x \leq 0,\\
(0.00; 0.25), & \quad \textrm{if}\, x \in (0; 1),\\
\{0.25\},     & \quad \textrm{if}\, x \in [1; 4],\\
(0.25; 0.50), & \quad \textrm{if}\, x \in (4; 4.5),\\
\{0.5\}       & \quad \textrm{if}\, x = 4.5,\\
(0.50; 0.75), & \quad \textrm{if}\, x \in (4.5; 5),\\
\{0.75\},     & \quad \textrm{if}\, x \in [5; 8],\\
(0.75; 1.00), & \quad \textrm{if}\, x \in (8; 9),\\
\{1.00\},     & \quad \textrm{if}\, x \geq 9.\\
\end{cases}
\]

Since the median value \(M=4.5\), we have:

\[
v_p \in \begin{cases}
[0.0; 0.5), & \quad \textrm{if}\, p < 0.5,\\
(3.5; 4.5), & \quad \textrm{if}\, p > 0.5.
\end{cases}
\]

In the point \(p = 0.5\), \(v_p\) jumps from \(0.5\) to \(3.5\),
which corresponds to a vertical segment in Figure~\ref{fig:qad-functions}d.
That is why \(\MAD(X)=\QAD(X, 0.5)\) is so unstable as shown in Figure~\ref{fig:mad-multimodal}b.

\hypertarget{poisson-distribution}{%
\subsubsection{Poisson distribution}\label{poisson-distribution}}

The CDF of the \(\Pois(\lambda)\) distribution is given by:

\[
F(x) = e^{-\lambda} \sum_{i=0}^{\lfloor x \rfloor} \frac{\lambda^i}{i!}.
\]

Unfortunately, there is no explicit analytic formula to express
the exact quantile values including the median
(see \autocite{adell2005,choi1994} for lower and upper bounds and approximations).
Therefore, we do not derive the exact equation for the \(\QAD\).
Instead of it, we just highlight the fact that the \(\QAD\)
of a discrete distribution is a piecewise linear function as shown in Figure~\ref{fig:qad-functions}e.
The biggest caveat for using such distributions is that the \(\MAD\) value can be zero so
we cannot use it as a denominator in various exactions.
In this case, if we want to continue using the \(\QAD\), we should consider higher values of \(p\).

\hypertarget{pareto-distribution}{%
\subsubsection{Pareto distribution}\label{pareto-distribution}}

We consider the \(\Pareto\) distribution
given by \(F(x)=1-1/x\) with the median value \(M=2\).
Since \(F\) is defined only for \(x \geq x_{\min} = 1\), we have to consider two cases: \(M - v_p \leq 1\) and \(M - v_p > 1\).
The critical value \(v_p^*\) is defined by \(v_p^* = M - x_{\min} = 1\).
Now it is easy to get the value of \(p^*\) using Equation~\eqref{eq:qad-pstar}:

\[
p^* = F(2M - x_{\min}) = F(3) = 2/3.
\]

Let us consider the first case when \(p \leq p^* = 2/3\).
Equation~\eqref{eq:qad-solve} has the following form:

\[
\Big( 1 - \frac{1}{2 + v_p} \Big) - \Big( 1 - \frac{1}{2 - v_p} \Big) = p,
\]

which is the same as

\[
\frac{1}{2 - v_p} - \frac{1}{2 + v_p} = p.
\]

By multiplying both sides of this equation on \((2 - v_p)(2 + v_p)\), we get:

\[
(2 + v_p) - (2 - v_p) = p (2 - v_p)(2 + v_p),
\]

which is the same as

\[
\frac{2}{p} v_p = 4 - v_p^2.
\]

Hence,

\[
v_p^2 + \frac{2}{p} v_p - 4 = 0.
\]

This is a quadratic equation for \(v_p\) with a solution given by

\[
v_p = \frac{-2/p \pm \sqrt{4/p^2+16}}{2} = \frac{-1}{p} \pm \sqrt{\frac{1}{p^2} + 4}.
\]

Since \(v_p\) is always positive, only the plus is applicable for \(\pm\).

Now let us consider the second case when \(p > p^* = 2/3\).
Equation~\eqref{eq:qad-solve2} has the following form:

\[
\Big( 1 - \frac{1}{2 + v_p} \Big) = p,
\]

which is the same as

\[
(2+v_p)(1 - p) = 1.
\]

Hence,

\[
2 - 2p + v_p - p v_p = 1.
\]

From this, we can express \(v_p\):

\[
v_p = \frac{2p - 1}{1 - p}.
\]

Thus, if \(X \sim \Pareto\),

\[
\lim_{n \to \infty} \E[\QAD(X, p)] = \begin{cases}
\frac{-1}{p} + \sqrt{\frac{1}{p^2} + 4}, & \textrm{if}\; p \leq 2/3,\\
\frac{2p - 1}{1 - p}, & \textrm{if}\; p > 2/3.
\end{cases}
\]

The corresponding plot is presented in Figure~\ref{fig:qad-functions}f.

\clearpage

\hypertarget{sec:qad-cc}{%
\subsection{Asymptotic consistency constants for QAD}\label{sec:qad-cc}}

Let \(\QAD_\infty(X, p)\) be an asymptotically consistent estimator
for the standard deviation under the normal distribution.
We define such an estimator as a product of \(\QAD(X, p)\) and a consistency constant \(K_p\):

\[
\QAD_\infty(X, p) = K_p \cdot \QAD(X, p) = K_p \cdot \Q(|X - \median(X)|, p).
\]

Let us assume that \(X\) follows the standard normal distribution \(\mathcal{N}(0, 1)\).
Since we want to achieve \(\lim_{n \to \infty} \E[\QAD_\infty(X, p)] = 1\), we have

\[
\lim_{n \to \infty} \E[\QAD(X, p)] = \frac{1}{K_p}.
\]

Using Equation~\eqref{eq:qad-normal}, we get the exact value for the asymptotic consistency constant value:

\begin{equation}
K_p = \dfrac{1}{\Phi^{-1}((p+1)/2)}.
\label{eq:kp}
\end{equation}

\hypertarget{sec:qad-age}{%
\subsection{Asymptotic Gaussian efficiency of QAD}\label{sec:qad-age}}

Let us consider \(X\) from the standard normal distribution: \(X \sim \mathcal{N}(0, 1)\).
For the normal model, \(\lim_{n \to \infty} \E[\Q(X, 0.5)] = 0\).
Therefore,

\[
\lim_{n \to \infty} \QAD_n(X, p) = K_p \cdot Q(|X|, p) = \frac{1}{\Phi^{-1}((p+1)/2)} \cdot Q(|X|, p).
\]

If \(X\) follows the standard normal distribution, \(|X|\) follows the standard half-normal distribution.
The probability density function and the quantile function
of the standard half-normal distribution are well-known and given by:

\[
f_{\operatorname{HN}}(x) = \sqrt{\frac{2}{\pi}} \operatorname{exp}(-x^2/2), \quad
Q_{\operatorname{HN}}(p) = \Phi^{-1}((p+1)/2).
\]

The asymptotic variance of the sample quantile estimator for distribution with probability density function \(f\)
and quantile function \(Q\) is defined as follows (see \autocite{wilcox2016}, 3.5.1):

\[
\lim_{n \to \infty} \V(Q_n(X, p)) = \dfrac{p(1-p)}{n f(Q(p))^2}.
\]

Using the definitions of \(f_{\operatorname{HN}}\) and \(Q_{\operatorname{HN}}\), we get

\[
\lim_{n \to \infty} \V(\QAD_n(X, p)) = \frac{1}{\big(\Phi^{-1}((p+1)/2)\big)^2} \cdot
 \frac{\pi p(1-p)}{2n} \cdot \exp\Big(\big(\Phi^{-1}((p+1)/2)\big)^2 \Big).
\]

The asymptotic variance of the standard deviation estimator is well-known:

\[
\lim_{n \to \infty} \V(\SD_n) = \frac{1}{2n}.
\]

Finally, we are ready to draw the equation for the asymptotic Gaussian efficiency of the \(\QAD\):

\begin{equation}
\begin{split}
\lim_{n \to \infty} e(\QAD_n(X, p),\; \SD_n(X)) =
  \lim_{n \to \infty} \frac{\V[\SD_n(X)]}{\V[\QAD_n(X, p)]} = \\
  = \Bigg( \frac{1}{\big(\Phi^{-1}((p+1)/2)\big)^2} \pi p(1-p) \exp\Big(\big(\Phi^{-1}((p+1)/2)\big)^2 \Big) \Bigg)^{-1} = \\
  = \frac{\big(\Phi^{-1}((p+1)/2)\big)^2}{\pi p(1-p) \exp\Big(\big(\Phi^{-1}((p+1)/2)\big)^2 \Big)}.
\end{split}
\label{eq:qad-asympt-eff}
\end{equation}

Using Equation~\eqref{eq:qad-asympt-eff}, we can obtain the exact value of the \(\MAD\) Gaussian efficiency
by putting \(p = 0.5\):

\[
\begin{split}
\lim_{n \to \infty} e(\MAD_n(X),\; \SD_n(X)) & \approx 0.367522937595603 \approx 36.75\%.
\end{split}
\]

\begin{figure}[ht!]

{\centering \includegraphics{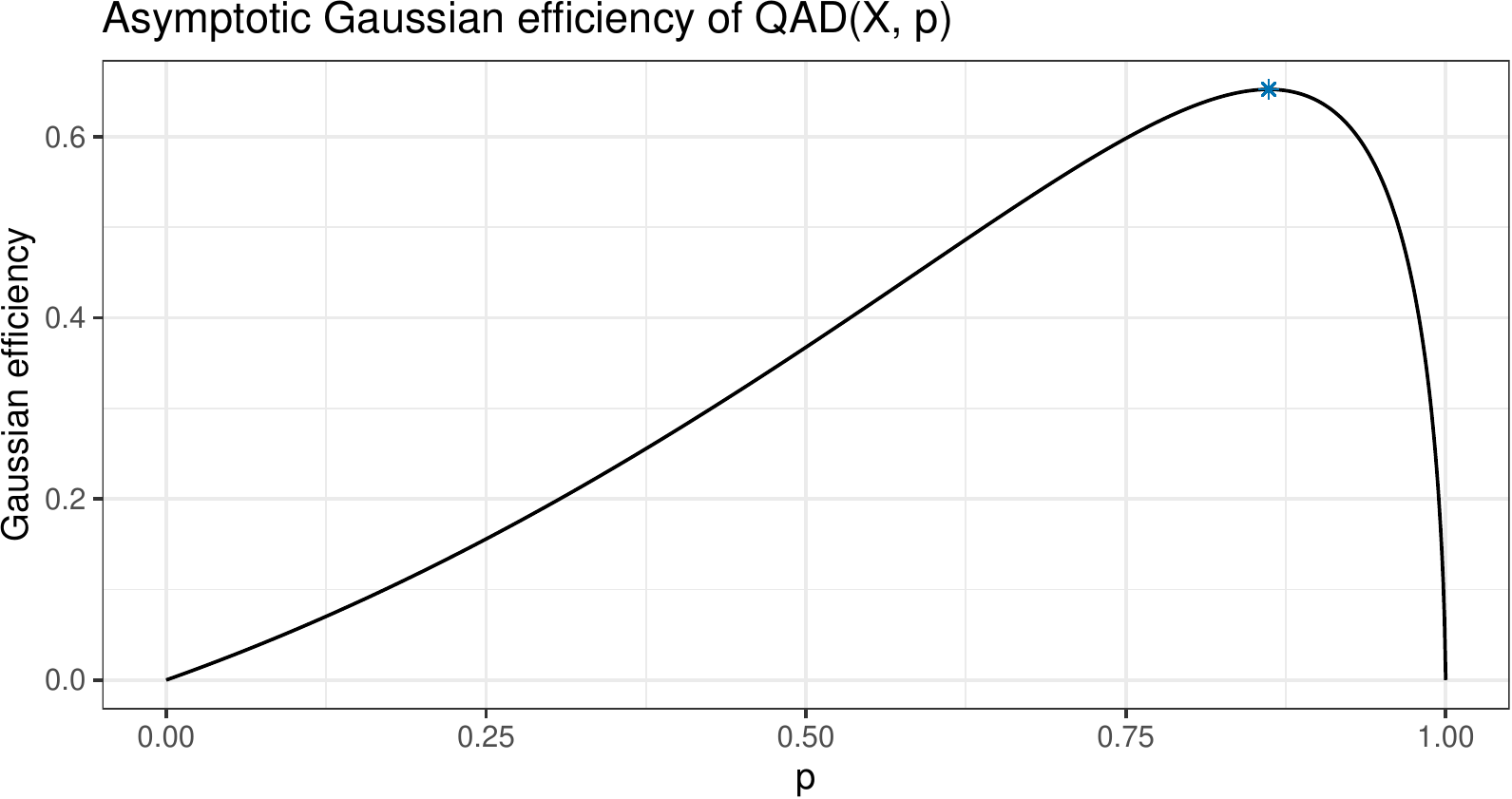} 

}

\caption{Asymptotic Gaussian efficiency of QAD(X, p).}\label{fig:qad-efficiency}
\end{figure}

The plot of the asymptotic Gaussian efficiency function for all values
\(p \in [0; 1]\) is shown in Figure~\ref{fig:qad-efficiency}.

We can see that the presented function is unimodal with a single maximum point.
Let us denote the location of this point as \(\rho_o\).
This value can be obtained numerically:

\begin{equation}
\rho_o \approx 0.861678977787423 \approx 86.17\%.
\label{eq:po}
\end{equation}

If we are looking for a specific value of \(p\) to get
a single measure of dispersion in order to use it as a robust estimator of the standard deviation,
it does not make sense to consider \(\QAD(X, p)\) for \(p > \rho_o\).
Indeed, such an estimator is worse than \(\QAD(X, \rho_o)\)
in terms of the Gaussian efficiency and breakdown point at the same time.
Meanwhile, it also does not make sense to consider \(p < 0.5\) since
\(50\%\) is the best possible breakdown point of a scale estimator (see \autocite[p14]{rousseeuw1987}).
Therefore, we consider only \(p \in [0.5; \rho_o]\).

\clearpage

\hypertarget{sec:qad-sqad}{%
\subsection{Standard quantile absolute deviation}\label{sec:qad-sqad}}

Manual observing of the \(\QAD(X, p)\) values for all \(p\) is not always convenient.
It is better to have a rule of thumb for choosing such a value of \(p\) that gives us desirable properties
of the corresponding measure of dispersion.
The \(\MAD\) approach suggests using \(p=0.5\) since it gives the highest breakdown point of \(50\%\).
However, such robustness is not always required in practice.
If the desired breakdown point is below \(50\%\), it is reasonable to increase the target value of \(p\) and
trade some robustness for statistical efficiency.

The first rule of thumb we want to suggest is using the following value of \(p\):

\[
\rho_s = \Phi(1) - \Phi(-1) \approx 0.682689492137086 \approx 68.27\%.
\]

The intuition behind this value is as follows.
If we assume only slight deviations from normality and obtain an estimation for the standard deviation \(\sigma\),
we may also expect that the interval \([\mu-\sigma; \mu+\sigma]\) is not so distorted
compared to the normal distribution.
The portion of the normal distribution covered by this interval is exactly \(\rho_s = \Phi(1) - \Phi(-1)\).
Therefore, the maximum asymptotic gross error percentage that does not corrupt \([\mu-\sigma; \mu+\sigma]\) is
\(1-\rho_s \approx 31.73\%\), which can be considered as the target breakdown point.

Let us denote the corresponding measure of dispersion by the \emph{standard quantile absolute deviation} or the \(\SQAD\).
Asymptotically, it can be defined as follows:

\[
\SQAD_\infty(X) = K_{\rho_s} \cdot \QAD(X, \rho_s),
\]

From \eqref{eq:kp}, the asymptotic consistency constant \(K_{\rho_s} = 1\) so
the \(\QAD(X, \rho_s)\) is an unbiased consistent estimator for the standard deviation under normality.
Although, for finite samples, we need bias-correction factors:

\[
\SQAD_n(X) = K_{s,n} \cdot \QAD(X, \rho_s),
\]

where \(K_{s,n}\) is a finite-sample consistency constant for the \(\SQAD\).
We obtain the \(K_{s,n}\) values in Section~\ref{sec:sim-constants}.

The asymptotic Gaussian efficiency of the \(\SQAD\) can be obtained using Equation~\eqref{eq:qad-asympt-eff}:

\[
\lim_{n \to \infty} e(\SQAD_n(X),\; \SD_n(X)) \approx 0.540565062173643 \approx 54.06\%.
\]

\hypertarget{sec:qad-oqad}{%
\subsection{Optimal quantile absolute deviation}\label{sec:qad-oqad}}

When the expected portion of gross errors is less than \(\rho_o \approx 86.17\%\) (given by Equation~\eqref{eq:po}),
it makes sense to use \(\QAD(X, \rho_o)\) since it gives the highest possible value of Gaussian efficiency.
Let us denote it by the \emph{optimal quantile absolute deviation} or the \(\OQAD\).
Here is the asymptotic expression for this measure:

\[
\OQAD_\infty(X) = K_{\rho_o} \cdot \QAD(X, \rho_o),
\]

From \eqref{eq:kp}, the asymptotic consistency constant \(K_{\rho_o} \approx 0.6747309\).
For finite samples, we need constant adjustments:

\[
\OQAD_n(X) = K_{o,n} \cdot \QAD(X, \rho_s),
\]

where \(K_{o,n}\) is a finite-sample consistency constant for the \(\OQAD\).
We obtain the \(K_{o,n}\) values in Section~\ref{sec:sim-constants}.

The asymptotic breakdown point of the \(\OQAD\) is \(1-\rho_o \approx 13.83\%\).
The asymptotic Gaussian efficiency of the \(\OQAD\) can be obtained using Equation~\eqref{eq:qad-asympt-eff}:

\[
\lim_{n \to \infty} e(\OQAD_n(X),\; \SD_n(X)) \approx 0.652244854073207 \approx 64.22\%.
\]

\clearpage

\hypertarget{sec:thdme}{%
\section{Trimmed Harrell-Davis median estimator}\label{sec:thdme}}

One of the straightforward insights that is provided by the \(\QAD\) is that the interval
\([\median(X) - \QAD(X, p); \median(X) + \QAD(X, p)]\) covers \(p \cdot 100\%\) of the distribution.
Since the asymptotic breakdown point of \(\QAD(X, p)\) is \(1-p\), it makes sense to consider a complimentary median estimator
which also has the asymptotic breakdown point of \(1-p\), but higher Gaussian efficiency.
We suggest using the trimmed Harrell-Davis quantile estimator based
on the highest density interval of the given width (see \autocite{akinshin2022thdqe}).

The classic Harrell-Davis quantile estimator (see \autocite{harrell1982}) is defined as follows:

\[
Q_{\operatorname{HD}}(X, q) = \sum_{i=1}^{n} W_{\operatorname{HD},i} \cdot X_{(i)},\quad
W_{\operatorname{HD},i} = I_{i/n}(\alpha, \beta) - I_{(i-1)/n}(\alpha, \beta),
\]

where \(I_u(\alpha, \beta)\) is the regularized incomplete beta function,
\(\alpha = (n+1)q\), \(\;\beta = (n+1)(1-q)\).
This approach suggests estimating quantiles as a weighted sum of order statistics.
While \(Q_{\operatorname{HD}}\) has higher Gaussian efficiency than the sample quantiles,
its breakdown point is zero.
However, most of the \(W_{\operatorname{HD},i}\) are negligible:
their actual impact on the Gaussian efficiency is low, but they significantly decrease the breakdown point.
When we switch to the trimmed modification of this estimator,
we perform summation only within the highest density interval \([L;R]\) of \(\operatorname{Beta}(\alpha, \beta)\)
of size \(D\):

\[
Q_{\operatorname{THD},D}(X, q) = \sum_{i=1}^{n} W_{\operatorname{THD},D,i} \cdot X_{(i)}, \quad
W_{\operatorname{THD},D,i} = F_{\operatorname{THD},D}(i / n) - F_{\operatorname{THD},D}((i - 1) / n),
\]

\[
F_{\operatorname{THD},D}(u) = \begin{cases}
0 & \textrm{for }\, u < L,\\
\big( I_u(\alpha, \beta) - I_L(\alpha, \beta) \big) /
\big( I_R(\alpha, \beta) \big) - I_L(\alpha, \beta) \big) \big)
  & \textrm{for }\, L \leq u \leq R,\\
1 & \textrm{for }\, R < u.
\end{cases}
\]

Thus, we use only sample elements with the highest weight coefficients \(W_{\operatorname{THD},D,i}\) and
ignore sample elements with small weight coefficients.
This allows us to get a customizable trade-off between the breakdown point and the Gaussian efficiency.
When we consider the median estimator \(Q_{\operatorname{THD},D}(0.5)\), we have \(\alpha=\beta=(n+1)/2\),
and the highest density interval \([L;R]\) is just \([0.5-p/2; 0.5+p/2]\).
Thus, we get the following form for the \emph{trimmed Harrell-Davis median estimator} \(\THDME_p\):

\[
\THDME_p(X) =
Q_{\operatorname{THD},p}(X, 0.5) = \sum_{i=1}^{n} W_{\operatorname{THDM},p,i} \cdot X_{(i)}, \quad
W_{\operatorname{THDM},p,i} = F_{\operatorname{THDM},p}(i / n) - F_{\operatorname{THDM},p}((i - 1) / n),
\]

\[
F_{\operatorname{THDM,p}}(u) = \begin{cases}
0 & \textrm{for }\, u < 0.5 - p/2,\\
\dfrac{ I_u(\frac{n+1}{2}, \frac{n+1}{2}) - I_{0.5-p/2}(\frac{n+1}{2}, \frac{n+1}{2}) }{I_{0.5+p/2}(\frac{n+1}{2}, \frac{n+1}{2}) - I_{0.5-p/2}(\frac{n+1}{2}, \frac{n+1}{2})}
  & \textrm{for }\, 0.5-p/2 \leq u \leq 0.5+p/2,\\
1 & \textrm{for }\, 0.5+p/2 < u.
\end{cases}
\]

For the \(\SQAD\), we consider the \emph{standard trimmed Harrell-Davis median estimator} (\(\STHDME\)) given by:

\[
\STHDME(X) = \THDME_{\rho_s}(X).
\]

For the \(\OQAD\), we consider the \emph{optimal trimmed Harrell-Davis median estimator} (\(\OTHDME\)) given by:

\[
\OTHDME(X) = \THDME_{\rho_o}(X).
\]

\clearpage

\hypertarget{sec:sim}{%
\section{Simulation studies}\label{sec:sim}}

In this section, we perform several simulations studies in order to obtain
the finite-sample consistency constants for the \(\SQAD\) and the \(\OQAD\) (Section~\ref{sec:sim-constants}),
the finite-sample Gaussian efficiency values of the \(\MAD\), the \(\SQAD\), and the \(\OQAD\) (Section~\ref{sec:sim-scale-efficiency}),
and the finite-sample Gaussian efficiency of the \(\STHDME\) and the \(\OTHDME\) (Section~\ref{sec:sim-location-efficiency}).

\hypertarget{sec:sim-constants}{%
\subsection{Simulation 1: The finite-sample consistency constants for the SQAD and the OQAD}\label{sec:sim-constants}}

Since \(K_{s,n} = 1/\E[\SQAD(X)]\) and \(K_{o,n} = 1/\E[\OQAD(X)]\),
the values of the finite-sample consistency constants for the \(\SQAD\) and the \(\OQAD\) can be obtained
by estimating the expected value of \(\SQAD(X)\) and \(\OQAD(X)\) using the Monte-Carlo method.
We do it according to the following scheme:

\begin{algorithm}[H]
\ForEach{$n \in \{ 2..100, \ldots, 10\,000 \}$}{
  $\textit{repetitions} \gets 25\,000\,000$\\
  \For{$i \gets 1..\textit{repetitions}$}{
        $x \gets \textrm{GenerateRandomSample}(\textrm{Distribution} = \mathcal{N}(0, 1),\, \textrm{SampleSize} = n)$\\
        $y_{\SQAD,i} \gets \textrm{SQAD}(x)$\\
        $y_{\OQAD,i} \gets \textrm{OQAD}(x)$
  }
  $K_{s,n} \gets 1 / (\sum y_{\SQAD,i} / \textit{repetitions})$\\
  $K_{o,n} \gets 1 / (\sum y_{\OQAD,i} / \textit{repetitions})$
}
\end{algorithm}

In order to estimate the \(\SQAD\) and the \(\OQAD\),
we use the sample median and the traditional Hyndman-Fan Type 7 quantile estimator
in order to make the presented results more generic.
If the \(\STHDME\) and the \(\OTHDME\) are used to estimate the median,
the consistency constant may require some adjustments.

\begin{figure}[ht!]

{\centering \includegraphics{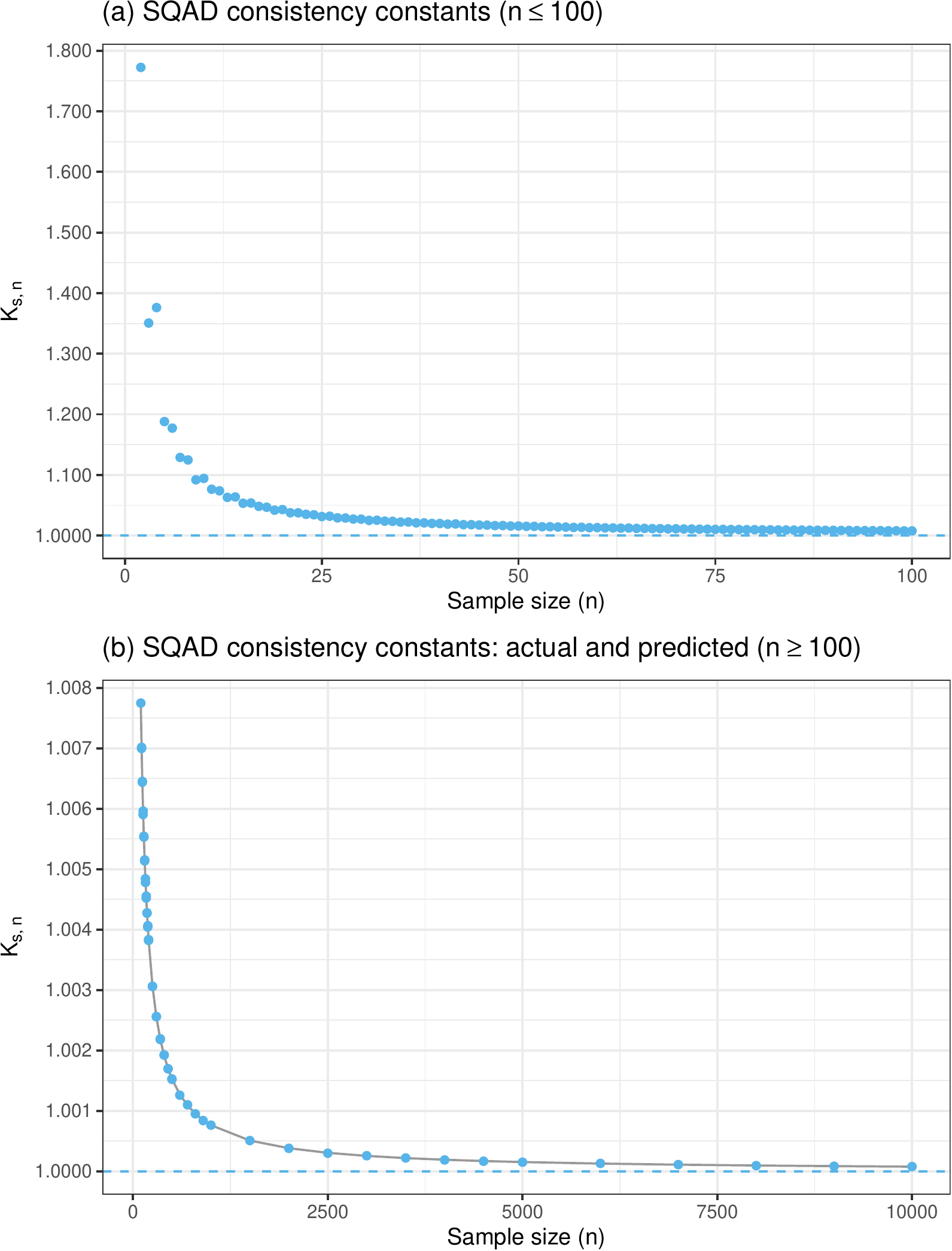} 

}

\caption{Finite-sample consistency constants for the SQAD.}\label{fig:fig-sqad-constants}
\end{figure}

\begin{figure}[ht!]

{\centering \includegraphics{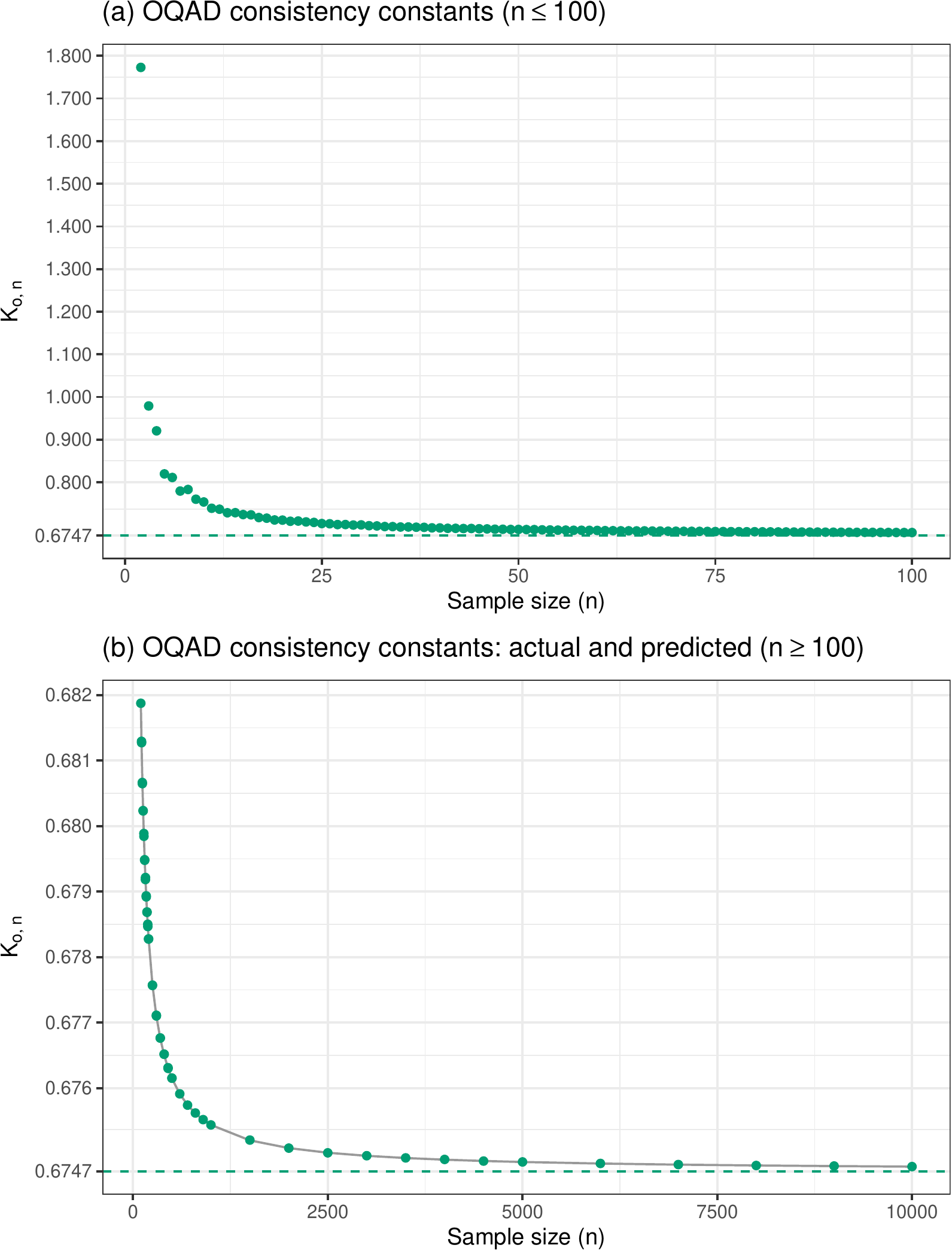} 

}

\caption{Finite-sample consistency constants for the OQAD.}\label{fig:fig-oqad-constants}
\end{figure}

The estimated \(K_{s,n},\, K_{o,n}\) values are presented in Table~\ref{tab:tab-constants}.
The corresponding plots for \(2 \leq n \leq 100\)
are shown in Figure~\ref{fig:fig-sqad-constants}a and Figure~\ref{fig:fig-oqad-constants}a.

Following the approach from \autocite{hayes2014},
we are going to draw generic \(K_n\) equations for \(n > 100\) in the following form:

\[
K_n = K_{\infty} (1 + \alpha n^{-1} + \beta n^{-2}),
\]

where \(K_{\infty}\) is the corresponding asymptotic consistency constant.

Using least squares on the values Table~\ref{tab:tab-constants} for \(100 \leq n \leq 1000\),
we can obtain approximated values of \(\alpha\) and \(\beta\),
which gives us the following equations for \(K_{s,n}\) (\(K_{\rho_s} = 1\) as shown in Section~\ref{sec:qad-sqad}):

\[
K_{s,n} \approx 1 +
  0.762 n^{-1} +
  0.967 n^{-2}.
\]

Using the same approach, we obtain the corresponding equation for \(K_{o,n}\)
(\(K_{\rho_o} \approx 0.6747309\) as shown in Section~\ref{sec:qad-oqad}):

\[
K_{o,n} \approx 0.6747309 \cdot (1 +
  1.047 n^{-1} +
  1.193 n^{-2}).
\]

The actual and predicted values of \(K_{s,n},\, K_{o,n}\)
for \(100 < n \leq 10\,000\)
are shown in Figure~\ref{fig:fig-sqad-constants}b and Figure~\ref{fig:fig-oqad-constants}b respectively.
The obtained values of \(\alpha\) and \(\beta\) look quite accurate:
the maximum observed absolute difference between the actual and predicted values
is \(\approx 0.000073\) for the \(\SQAD\)
and \(\approx 0.00006\) for the \(\OQAD\).

\clearpage

\begin{table}[!h]

\caption{\label{tab:tab-constants}Finite-sample consistency constants for the $\SQAD$ and the $\OQAD$.}
\centering
\begin{tabular}[t]{r|r|r|r|r|r|r|r|r}
\hline
n & $\SQAD$ & $\OQAD$ & n & $\SQAD$ & $\OQAD$ & n & $\SQAD$ & $\OQAD$\\
\hline
1 & - & - & 51 & 1.0152 & 0.6888 & 109 & 1.0070 & 0.6813\\
\hline
2 & 1.7724 & 1.7729 & 52 & 1.0152 & 0.6887 & 110 & 1.0070 & 0.6813\\
\hline
3 & 1.3506 & 0.9788 & 53 & 1.0146 & 0.6882 & 119 & 1.0064 & 0.6807\\
\hline
4 & 1.3762 & 0.9205 & 54 & 1.0146 & 0.6880 & 120 & 1.0065 & 0.6806\\
\hline
5 & 1.1881 & 0.8194 & 55 & 1.0141 & 0.6875 & 129 & 1.0059 & 0.6802\\
\hline
6 & 1.1773 & 0.8110 & 56 & 1.0140 & 0.6875 & 130 & 1.0060 & 0.6802\\
\hline
7 & 1.1289 & 0.7792 & 57 & 1.0135 & 0.6871 & 139 & 1.0055 & 0.6799\\
\hline
8 & 1.1248 & 0.7828 & 58 & 1.0137 & 0.6872 & 140 & 1.0055 & 0.6798\\
\hline
9 & 1.0920 & 0.7600 & 59 & 1.0130 & 0.6870 & 149 & 1.0051 & 0.6795\\
\hline
10 & 1.0943 & 0.7535 & 60 & 1.0131 & 0.6868 & 150 & 1.0052 & 0.6795\\
\hline
11 & 1.0764 & 0.7388 & 61 & 1.0127 & 0.6863 & 159 & 1.0048 & 0.6792\\
\hline
12 & 1.0738 & 0.7365 & 62 & 1.0126 & 0.6862 & 160 & 1.0048 & 0.6792\\
\hline
13 & 1.0630 & 0.7282 & 63 & 1.0123 & 0.6859 & 169 & 1.0045 & 0.6789\\
\hline
14 & 1.0637 & 0.7284 & 64 & 1.0124 & 0.6859 & 170 & 1.0046 & 0.6789\\
\hline
15 & 1.0533 & 0.7241 & 65 & 1.0118 & 0.6857 & 179 & 1.0043 & 0.6787\\
\hline
16 & 1.0537 & 0.7234 & 66 & 1.0119 & 0.6858 & 180 & 1.0043 & 0.6787\\
\hline
17 & 1.0482 & 0.7170 & 67 & 1.0115 & 0.6854 & 189 & 1.0040 & 0.6785\\
\hline
18 & 1.0468 & 0.7155 & 68 & 1.0115 & 0.6853 & 190 & 1.0041 & 0.6785\\
\hline
19 & 1.0419 & 0.7113 & 69 & 1.0111 & 0.6850 & 199 & 1.0038 & 0.6783\\
\hline
20 & 1.0429 & 0.7110 & 70 & 1.0112 & 0.6849 & 200 & 1.0038 & 0.6783\\
\hline
21 & 1.0377 & 0.7083 & 71 & 1.0108 & 0.6847 & 249 & 1.0031 & 0.6776\\
\hline
22 & 1.0376 & 0.7088 & 72 & 1.0108 & 0.6847 & 250 & 1.0031 & 0.6776\\
\hline
23 & 1.0351 & 0.7068 & 73 & 1.0106 & 0.6846 & 299 & 1.0026 & 0.6771\\
\hline
24 & 1.0343 & 0.7056 & 74 & 1.0106 & 0.6845 & 300 & 1.0026 & 0.6771\\
\hline
25 & 1.0314 & 0.7030 & 75 & 1.0102 & 0.6842 & 349 & 1.0022 & 0.6768\\
\hline
26 & 1.0320 & 0.7024 & 76 & 1.0103 & 0.6841 & 350 & 1.0022 & 0.6768\\
\hline
27 & 1.0292 & 0.7006 & 77 & 1.0100 & 0.6839 & 399 & 1.0019 & 0.6765\\
\hline
28 & 1.0290 & 0.7006 & 78 & 1.0100 & 0.6839 & 400 & 1.0019 & 0.6765\\
\hline
29 & 1.0272 & 0.6995 & 79 & 1.0097 & 0.6837 & 449 & 1.0017 & 0.6763\\
\hline
30 & 1.0271 & 0.6998 & 80 & 1.0097 & 0.6838 & 450 & 1.0017 & 0.6763\\
\hline
31 & 1.0251 & 0.6979 & 81 & 1.0095 & 0.6836 & 499 & 1.0015 & 0.6762\\
\hline
32 & 1.0253 & 0.6974 & 82 & 1.0095 & 0.6834 & 500 & 1.0015 & 0.6762\\
\hline
33 & 1.0238 & 0.6960 & 83 & 1.0093 & 0.6833 & 600 & 1.0013 & 0.6759\\
\hline
34 & 1.0235 & 0.6958 & 84 & 1.0092 & 0.6832 & 700 & 1.0011 & 0.6757\\
\hline
35 & 1.0223 & 0.6949 & 85 & 1.0090 & 0.6831 & 800 & 1.0010 & 0.6756\\
\hline
36 & 1.0224 & 0.6949 & 86 & 1.0091 & 0.6830 & 900 & 1.0008 & 0.6755\\
\hline
37 & 1.0210 & 0.6944 & 87 & 1.0089 & 0.6829 & 1000 & 1.0008 & 0.6754\\
\hline
38 & 1.0210 & 0.6940 & 88 & 1.0088 & 0.6830 & 1500 & 1.0005 & 0.6752\\
\hline
39 & 1.0201 & 0.6929 & 89 & 1.0086 & 0.6827 & 2000 & 1.0004 & 0.6751\\
\hline
40 & 1.0199 & 0.6927 & 90 & 1.0086 & 0.6827 & 2500 & 1.0003 & 0.6750\\
\hline
41 & 1.0189 & 0.6918 & 91 & 1.0084 & 0.6825 & 3000 & 1.0003 & 0.6750\\
\hline
42 & 1.0192 & 0.6918 & 92 & 1.0084 & 0.6825 & 3500 & 1.0002 & 0.6749\\
\hline
43 & 1.0180 & 0.6913 & 93 & 1.0082 & 0.6823 & 4000 & 1.0002 & 0.6749\\
\hline
44 & 1.0180 & 0.6914 & 94 & 1.0082 & 0.6823 & 4500 & 1.0002 & 0.6749\\
\hline
45 & 1.0174 & 0.6907 & 95 & 1.0081 & 0.6823 & 5000 & 1.0002 & 0.6749\\
\hline
46 & 1.0172 & 0.6904 & 96 & 1.0081 & 0.6822 & 6000 & 1.0001 & 0.6748\\
\hline
47 & 1.0165 & 0.6897 & 97 & 1.0079 & 0.6820 & 7000 & 1.0001 & 0.6748\\
\hline
48 & 1.0166 & 0.6896 & 98 & 1.0079 & 0.6820 & 8000 & 1.0001 & 0.6748\\
\hline
49 & 1.0158 & 0.6891 & 99 & 1.0078 & 0.6819 & 9000 & 1.0001 & 0.6748\\
\hline
50 & 1.0158 & 0.6892 & 100 & 1.0077 & 0.6819 & 10000 & 1.0001 & 0.6748\\
\hline
\end{tabular}
\end{table}

\clearpage

\clearpage

\clearpage

\hypertarget{sec:sim-scale-efficiency}{%
\subsection{Simulation 2: The finite-sample Gaussian efficiency values for the MAD, the SQAD, and the OQAD}\label{sec:sim-scale-efficiency}}

In this simulation study, we evaluate the finite-sample Gaussian efficiency of \(\MAD\), \(\SQAD\), and \(\OQAD\).
As for the baseline, we consider the unbiased standard deviation \(\SD_n\) of the normal distribution:

\[
\SD_n(X) = \sqrt{\frac{1}{n} \sum_{i=1}^n (X_i - \bar{X})^2} \bigg/ c_4(n), \quad
c_4(n) = \sqrt{\frac{2}{n-1}}\frac{\Gamma(\frac{n}{2})}{\Gamma(\frac{n-1}{2})}.
\]

For an unbiased scale estimator \(T_n\), the Gaussian efficiency is defined as follows:

\begin{equation}
e(T_n) = \frac{\V[\SD_n]}{\V[T_n]}.
\label{eq:dispersion-efficiency}
\end{equation}

We also consider the concept of the \emph{standardized asymptotic variance} of a scale estimator which was
proposed in \autocite{daniell1920} and advocated in \autocite[p1276]{rousseeuw1993}, \autocite[p502]{bickel1976}, and \autocite[p3]{huber2009}:

\begin{equation}
\V_s[T_n] = \frac{n \cdot \V[T_n]}{\E[T_n]^2}
\label{eq:svar}
\end{equation}

We suggest overriding Equation~\eqref{eq:dispersion-efficiency} using \(\V_s\):

\begin{equation}
e(T_n) = \frac{\V_s[\SD_n]}{\V_s[T_n]}.
\label{eq:dispersion-efficiency2}
\end{equation}

Equations~\eqref{eq:dispersion-efficiency} and \eqref{eq:dispersion-efficiency2} are equivalent for unbiased estimators
since \(\E[T_n] = 1\).
However, we operate only with approximations of the consistency constants
for the \(\MAD\) (see \autocite{park2020}), the \(\SQAD\) and the \(\OQAD\) (see Table~\ref{tab:tab-constants}).
The difference between the actual and approximated values
of the consistency constants is almost negligible in practice,
but it still introduces minor errors.
In order to slightly improve the accuracy of the estimated Gaussian efficiency values, we prefer using
Equation~\eqref{eq:dispersion-efficiency2} in our calculations.

We perform the simulation using the Monte-Carlo method according to the following scheme:

\begin{algorithm}[H]
\ForEach{$n \in \{  2..100, \ldots, 100\,000 \}$}{
  $\textit{repetitions} \gets 10\,000\,000$\\
  \For{$i \gets 1..\textit{repetitions}$}{
    $x \gets \textrm{GenerateRandomSample}(\textrm{Distribution} = \mathcal{N}(0, 1),\, \textrm{SampleSize} = n)$\\
    $y_{\SD,i} = \SD_n(x)$\\
    $y_{\MAD,i} = \MAD_n(x)$\\
    $y_{\SQAD,i} = \SQAD_n(x)$\\
    $y_{\OQAD,i} = \OQAD_n(x)$\\
  }
  $e(\MAD_n) = \V_s(y_{\SD,\{i\}}) / \V_s(y_{\MAD,\{i\}})$\\
  $e(\SQAD_n) = \V_s(y_{\SD,\{i\}}) / \V_s(y_{\SQAD,\{i\}})$\\
  $e(\OQAD_n) = \V_s(y_{\SD,\{i\}}) / \V_s(y_{\OQAD,\{i\}})$\\
}
\end{algorithm}

\begin{figure}[ht!]

{\centering \includegraphics{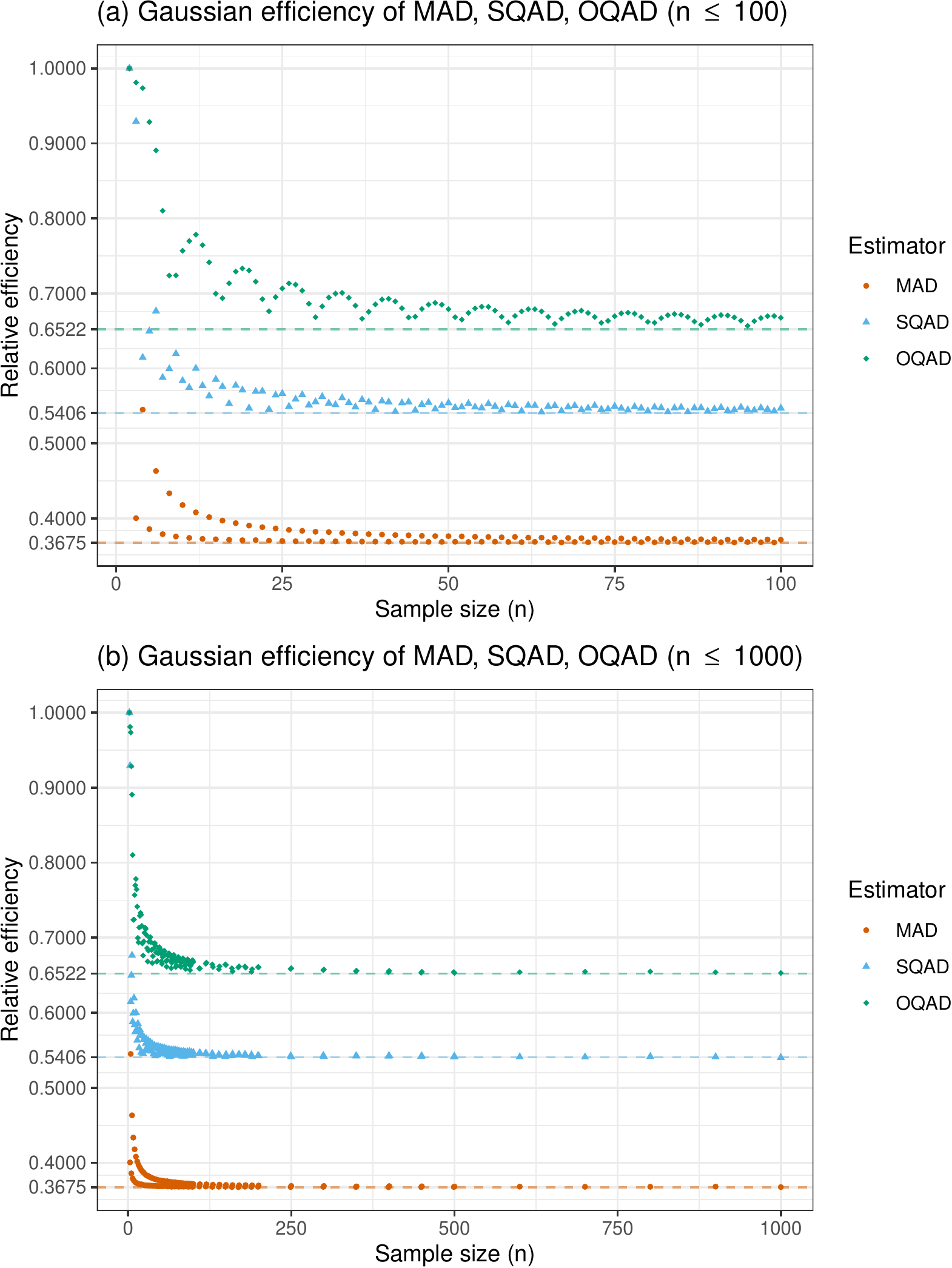} 

}

\caption{Finite-sample Gaussian efficiency of MAD, SQAD, OQAD.}\label{fig:fig-scale-efficiency}
\end{figure}

The estimated Gaussian efficiency values are presented in Table~\ref{tab:tab-scale-efficiency}.
The corresponding plots for \(n \leq 100\) and \(n \leq 1000\) are shown in Figure~\ref{fig:fig-scale-efficiency}.

As we can see, the \(\SQAD\) and the \(\OQAD\) are not only asymptotically more efficient than \(\MAD\),
but they are also more efficient for finite samples.
Thus, if the breakdown point of \(31.73\%\) is acceptable,
the \(\SQAD\) is preferable to the \(\MAD\).
If the breakdown point of \(13.83\%\) is acceptable,
the \(\OQAD\) is preferable to the \(\MAD\) and the \(\SQAD\)
since it gives the highest Gaussian efficiency among all \(\QAD(X, p)\) estimators.
For a custom breakdown point, one can choose an appropriate value of \(p\)
and perform a similar simulation in order to obtain the corresponding consistency constants for \(\QAD(X, p)\).

\clearpage

\begin{table}[!h]

\caption{\label{tab:tab-scale-efficiency}Finite-sample Gaussian efficiency of MAD, SQAD, OQAD.}
\centering
\begin{tabular}[t]{r|r|r|r|r|r|r|r|r|r|r|r}
\hline
\begingroup\fontsize{6}{8}\selectfont n\endgroup & \begingroup\fontsize{6}{8}\selectfont MAD\endgroup & \begingroup\fontsize{6}{8}\selectfont SQAD\endgroup & \begingroup\fontsize{6}{8}\selectfont OQAD\endgroup & \begingroup\fontsize{6}{8}\selectfont n\endgroup & \begingroup\fontsize{6}{8}\selectfont MAD\endgroup & \begingroup\fontsize{6}{8}\selectfont SQAD\endgroup & \begingroup\fontsize{6}{8}\selectfont OQAD\endgroup & \begingroup\fontsize{6}{8}\selectfont n\endgroup & \begingroup\fontsize{6}{8}\selectfont MAD\endgroup & \begingroup\fontsize{6}{8}\selectfont SQAD\endgroup & \begingroup\fontsize{6}{8}\selectfont OQAD\endgroup\\
\hline
1 & - & - & - & 51 & 0.3690 & 0.5479 & 0.6684 & 109 & 0.3684 & 0.5457 & 0.6591\\
\hline
2 & 1.0000 & 1.0000 & 1.0000 & 52 & 0.3761 & 0.5486 & 0.6651 & 110 & 0.3715 & 0.5467 & 0.6591\\
\hline
3 & 0.4005 & 0.9293 & 0.9810 & 53 & 0.3687 & 0.5532 & 0.6740 & 119 & 0.3683 & 0.5461 & 0.6653\\
\hline
4 & 0.5451 & 0.6146 & 0.9736 & 54 & 0.3756 & 0.5485 & 0.6801 & 120 & 0.3713 & 0.5450 & 0.6668\\
\hline
5 & 0.3859 & 0.6498 & 0.9285 & 55 & 0.3690 & 0.5473 & 0.6824 & 129 & 0.3680 & 0.5456 & 0.6644\\
\hline
6 & 0.4633 & 0.6765 & 0.8907 & 56 & 0.3754 & 0.5528 & 0.6822 & 130 & 0.3708 & 0.5418 & 0.6609\\
\hline
7 & 0.3791 & 0.5878 & 0.8102 & 57 & 0.3691 & 0.5496 & 0.6771 & 139 & 0.3680 & 0.5443 & 0.6580\\
\hline
8 & 0.4336 & 0.5994 & 0.7238 & 58 & 0.3749 & 0.5447 & 0.6689 & 140 & 0.3709 & 0.5431 & 0.6618\\
\hline
9 & 0.3760 & 0.6196 & 0.7240 & 59 & 0.3688 & 0.5515 & 0.6613 & 149 & 0.3680 & 0.5418 & 0.6643\\
\hline
10 & 0.4180 & 0.5836 & 0.7569 & 60 & 0.3754 & 0.5503 & 0.6699 & 150 & 0.3705 & 0.5445 & 0.6636\\
\hline
11 & 0.3741 & 0.5745 & 0.7698 & 61 & 0.3685 & 0.5432 & 0.6761 & 159 & 0.3677 & 0.5426 & 0.6592\\
\hline
12 & 0.4082 & 0.6000 & 0.7783 & 62 & 0.3746 & 0.5508 & 0.6790 & 160 & 0.3702 & 0.5450 & 0.6551\\
\hline
13 & 0.3728 & 0.5771 & 0.7643 & 63 & 0.3689 & 0.5506 & 0.6790 & 169 & 0.3675 & 0.5436 & 0.6604\\
\hline
14 & 0.4018 & 0.5633 & 0.7415 & 64 & 0.3745 & 0.5418 & 0.6760 & 170 & 0.3700 & 0.5445 & 0.6619\\
\hline
15 & 0.3723 & 0.5854 & 0.6997 & 65 & 0.3685 & 0.5493 & 0.6687 & 179 & 0.3681 & 0.5443 & 0.6617\\
\hline
16 & 0.3972 & 0.5758 & 0.6936 & 66 & 0.3743 & 0.5504 & 0.6594 & 180 & 0.3702 & 0.5435 & 0.6600\\
\hline
17 & 0.3716 & 0.5530 & 0.7136 & 67 & 0.3681 & 0.5434 & 0.6665 & 189 & 0.3678 & 0.5438 & 0.6544\\
\hline
18 & 0.3938 & 0.5773 & 0.7292 & 68 & 0.3740 & 0.5486 & 0.6726 & 190 & 0.3700 & 0.5415 & 0.6581\\
\hline
19 & 0.3710 & 0.5712 & 0.7332 & 69 & 0.3687 & 0.5504 & 0.6758 & 199 & 0.3680 & 0.5429 & 0.6609\\
\hline
20 & 0.3905 & 0.5471 & 0.7306 & 70 & 0.3738 & 0.5450 & 0.6771 & 200 & 0.3697 & 0.5424 & 0.6606\\
\hline
21 & 0.3712 & 0.5694 & 0.7157 & 71 & 0.3684 & 0.5471 & 0.6740 & 249 & 0.3677 & 0.5425 & 0.6585\\
\hline
22 & 0.3884 & 0.5697 & 0.6924 & 72 & 0.3739 & 0.5505 & 0.6685 & 250 & 0.3692 & 0.5412 & 0.6592\\
\hline
23 & 0.3705 & 0.5456 & 0.6762 & 73 & 0.3685 & 0.5454 & 0.6608 & 299 & 0.3679 & 0.5430 & 0.6567\\
\hline
24 & 0.3867 & 0.5645 & 0.6951 & 74 & 0.3734 & 0.5458 & 0.6634 & 300 & 0.3691 & 0.5417 & 0.6576\\
\hline
25 & 0.3701 & 0.5665 & 0.7066 & 75 & 0.3686 & 0.5496 & 0.6696 & 349 & 0.3677 & 0.5423 & 0.6554\\
\hline
26 & 0.3848 & 0.5493 & 0.7134 & 76 & 0.3734 & 0.5469 & 0.6734 & 350 & 0.3685 & 0.5422 & 0.6559\\
\hline
27 & 0.3697 & 0.5589 & 0.7117 & 77 & 0.3682 & 0.5444 & 0.6743 & 399 & 0.3675 & 0.5419 & 0.6540\\
\hline
28 & 0.3835 & 0.5648 & 0.7037 & 78 & 0.3729 & 0.5490 & 0.6730 & 400 & 0.3689 & 0.5426 & 0.6558\\
\hline
29 & 0.3700 & 0.5512 & 0.6862 & 79 & 0.3684 & 0.5472 & 0.6687 & 449 & 0.3675 & 0.5417 & 0.6532\\
\hline
30 & 0.3823 & 0.5554 & 0.6683 & 80 & 0.3729 & 0.5433 & 0.6621 & 450 & 0.3687 & 0.5424 & 0.6546\\
\hline
31 & 0.3695 & 0.5625 & 0.6828 & 81 & 0.3683 & 0.5484 & 0.6610 & 499 & 0.3677 & 0.5410 & 0.6542\\
\hline
32 & 0.3819 & 0.5536 & 0.6945 & 82 & 0.3726 & 0.5475 & 0.6671 & 500 & 0.3683 & 0.5418 & 0.6531\\
\hline
33 & 0.3696 & 0.5514 & 0.6998 & 83 & 0.3685 & 0.5420 & 0.6710 & 600 & 0.3682 & 0.5412 & 0.6541\\
\hline
34 & 0.3807 & 0.5607 & 0.7007 & 84 & 0.3728 & 0.5473 & 0.6725 & 700 & 0.3680 & 0.5409 & 0.6546\\
\hline
35 & 0.3695 & 0.5543 & 0.6942 & 85 & 0.3682 & 0.5482 & 0.6716 & 800 & 0.3680 & 0.5417 & 0.6548\\
\hline
36 & 0.3800 & 0.5486 & 0.6838 & 86 & 0.3724 & 0.5421 & 0.6686 & 900 & 0.3682 & 0.5412 & 0.6540\\
\hline
37 & 0.3693 & 0.5584 & 0.6662 & 87 & 0.3685 & 0.5473 & 0.6626 & 1000 & 0.3676 & 0.5402 & 0.6530\\
\hline
38 & 0.3792 & 0.5551 & 0.6755 & 88 & 0.3723 & 0.5476 & 0.6584 & 1500 & 0.3679 & 0.5413 & 0.6537\\
\hline
39 & 0.3692 & 0.5452 & 0.6854 & 89 & 0.3685 & 0.5430 & 0.6649 & 2000 & 0.3677 & 0.5407 & 0.6529\\
\hline
40 & 0.3786 & 0.5564 & 0.6919 & 90 & 0.3725 & 0.5460 & 0.6690 & 3000 & 0.3675 & 0.5407 & 0.6524\\
\hline
41 & 0.3691 & 0.5553 & 0.6929 & 91 & 0.3684 & 0.5479 & 0.6711 & 4000 & 0.3677 & 0.5408 & 0.6525\\
\hline
42 & 0.3779 & 0.5424 & 0.6894 & 92 & 0.3718 & 0.5438 & 0.6706 & 5000 & 0.3675 & 0.5405 & 0.6526\\
\hline
43 & 0.3690 & 0.5540 & 0.6803 & 93 & 0.3681 & 0.5448 & 0.6680 & 6000 & 0.3678 & 0.5408 & 0.6523\\
\hline
44 & 0.3775 & 0.5552 & 0.6680 & 94 & 0.3723 & 0.5481 & 0.6637 & 7000 & 0.3676 & 0.5407 & 0.6524\\
\hline
45 & 0.3688 & 0.5442 & 0.6690 & 95 & 0.3686 & 0.5451 & 0.6568 & 8000 & 0.3674 & 0.5406 & 0.6523\\
\hline
46 & 0.3769 & 0.5520 & 0.6792 & 96 & 0.3719 & 0.5443 & 0.6630 & 9000 & 0.3673 & 0.5406 & 0.6525\\
\hline
47 & 0.3689 & 0.5546 & 0.6852 & 97 & 0.3683 & 0.5474 & 0.6672 & 10000 & 0.3674 & 0.5402 & 0.6523\\
\hline
48 & 0.3766 & 0.5459 & 0.6874 & 98 & 0.3720 & 0.5455 & 0.6696 & 25000 & 0.3675 & 0.5407 & 0.6521\\
\hline
49 & 0.3688 & 0.5504 & 0.6849 & 99 & 0.3681 & 0.5432 & 0.6699 & 50000 & 0.3673 & 0.5405 & 0.6526\\
\hline
50 & 0.3765 & 0.5540 & 0.6790 & 100 & 0.3716 & 0.5470 & 0.6676 & 100000 & 0.3678 & 0.5409 & 0.6521\\
\hline
\end{tabular}
\end{table}

\clearpage

\clearpage

\hypertarget{sec:sim-location-efficiency}{%
\subsection{Simulation 3: The finite-sample Gaussian efficiency of the standard trimmed Harrell-Davis median estimator}\label{sec:sim-location-efficiency}}

In this simulation study, we evaluate the relative statistical efficiency
of the sample median (\(SM_n\)),
the standard and optimal trimmed Harrell-Davis median estimators (\(\STHDME_n\), \(\OTHDME_n\))
to the mean (\(\mean_n\)) under the normal distribution.
We call it the Gaussian efficiency of a location estimator or just the Gaussian efficiency
which will not introduce any confusion with the Gaussian efficiency of scale estimators.

The \(\SM_n\) is defined the traditional way as the middle order statistic or the sum of two middle order statistics
depending on the parity of \(n\):

\[
\SM_n(X) = \begin{cases}
X_{((n+1)/2)}, \quad & \textrm{if}\; n\; \textrm{is odd},\\
\dfrac{X_{(n/2)} + X_{(n/2+1)}}{2}, \quad & \textrm{if}\; n\; \textrm{is even}.
\end{cases}
\]

The \(\mean_n\) is just the arithmetic average:

\[
\mean_n(X) = \frac{X_1 + X_2 + \ldots + X_n}{n}.
\]

The definitions of \(\STHDME_n\) and \(\OTHDME_n\) are taken from Section~\ref{sec:thdme}.

We evaluate the Gaussian efficiency of an unbiased location estimator \(T_n\)
as a ratio between variances of \(\mean_n\) (the baseline) and \(T_n\):

\begin{equation}
e(T_n) = \frac{\V[\mean_n]}{\V[T_n]}.
\end{equation}

We perform the simulation using the Monte-Carlo method according to the following scheme:

\begin{algorithm}[H]
\ForEach{$n \in \{  2..100, \ldots, 100\,000 \}$}{
  $\textit{repetitions} \gets 10\,000\,000$\\
  \For{$i \gets 1..\textit{repetitions}$}{
    $x \gets \textrm{GenerateRandomSample}(\textrm{Distribution} = \mathcal{N}(0, 1),\, \textrm{SampleSize} = n)$\\
    $y_{\mean,i} = \mean_n(x)$\\
    $y_{\SM,i} = \SM_n(x)$\\
    $y_{\STHDME,i} = \STHDME_n(x)$\\
    $y_{\OTHDME,i} = \OTHDME_n(x)$\\
  }
  $e(\SM_n) = \V(y_{\mean,\{i\}}) / \V(y_{\SM,\{i\}})$\\
  $e(\STHDME_n) = \V(y_{\mean,\{i\}}) / \V(y_{\STHDME,\{i\}})$\\
  $e(\OTHDME_n) = \V(y_{\mean,\{i\}}) / \V(y_{\OTHDME,\{i\}})$\\
}
\end{algorithm}

\begin{figure}[ht!]

{\centering \includegraphics{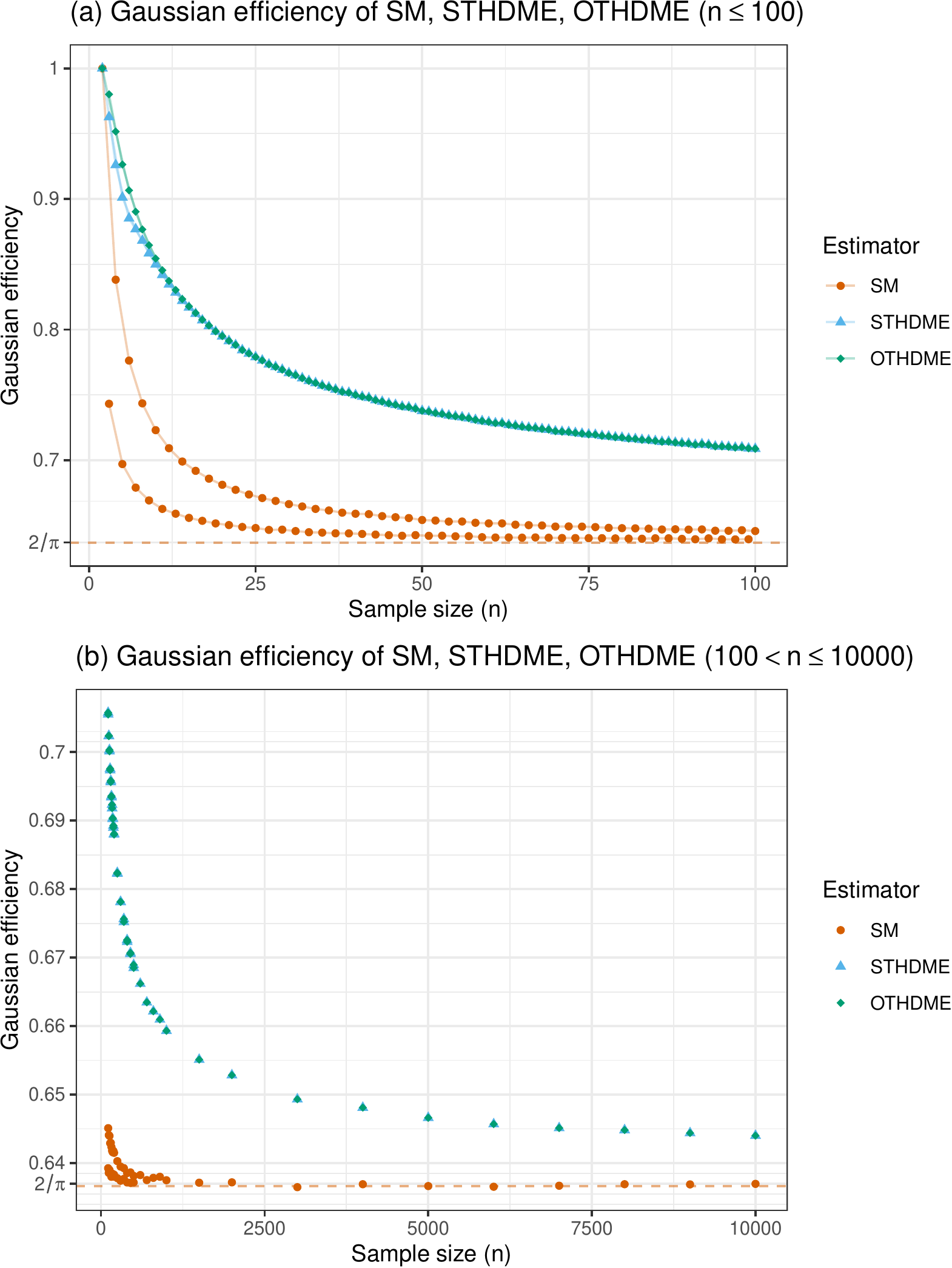} 

}

\caption{Finite-sample Gaussian efficiency of SM, STHDME, OTHDME.}\label{fig:fig-location-efficiency}
\end{figure}

The estimated Gaussian efficiency values are presented in Table~\ref{tab:tab-location-efficiency}.
The corresponding plots for \(n \leq 100\) and \(100 < n \leq 10\,000\)
are shown in Figure~\ref{fig:fig-location-efficiency}.

As we can see, the \(\STHDME_n\) and the \(\OTHDME_n\) are noticeably more efficient than the \(\SM\).
This makes the \(\STHDME_n\) and \(\OTHDME_n\) preferable option compared to \(\SM_n\)
when it is used together with the \(\SQAD_n\) and the \(\OQAD_n\) respectively
because they have identical breakdown points.

It is also worth mentioning that the finite-sample Gaussian efficiency values of the \(\STHDME_n\) and the \(\OTHDME_n\)
are almost equal, the difference is negligible (especially for \(n > 20\)).

\clearpage

\begin{table}[!h]

\caption{\label{tab:tab-location-efficiency}Finite-sample Gaussian efficiency of SM, STHDME, OTHDME.}
\centering
\begin{tabular}[t]{r|r|r|r|r|r|r|r|r|r|r|r}
\hline
\begingroup\fontsize{6}{8}\selectfont n\endgroup & \begingroup\fontsize{6}{8}\selectfont SM\endgroup & \begingroup\fontsize{6}{8}\selectfont STHDME\endgroup & \begingroup\fontsize{6}{8}\selectfont OTHDME\endgroup & \begingroup\fontsize{6}{8}\selectfont n\endgroup & \begingroup\fontsize{6}{8}\selectfont SM\endgroup & \begingroup\fontsize{6}{8}\selectfont STHDME\endgroup & \begingroup\fontsize{6}{8}\selectfont OTHDME\endgroup & \begingroup\fontsize{6}{8}\selectfont n\endgroup & \begingroup\fontsize{6}{8}\selectfont SM\endgroup & \begingroup\fontsize{6}{8}\selectfont STHDME\endgroup & \begingroup\fontsize{6}{8}\selectfont OTHDME\endgroup\\
\hline
1 & - & - & - & 51 & 0.6420 & 0.7371 & 0.7371 & 109 & 0.6392 & 0.7058 & 0.7058\\
\hline
2 & 1.0000 & 1.0000 & 1.0000 & 52 & 0.6537 & 0.7360 & 0.7360 & 110 & 0.6451 & 0.7055 & 0.7055\\
\hline
3 & 0.7431 & 0.9627 & 0.9801 & 53 & 0.6421 & 0.7353 & 0.7353 & 119 & 0.6386 & 0.7023 & 0.7023\\
\hline
4 & 0.8381 & 0.9260 & 0.9516 & 54 & 0.6529 & 0.7342 & 0.7342 & 120 & 0.6441 & 0.7024 & 0.7024\\
\hline
5 & 0.6970 & 0.9011 & 0.9264 & 55 & 0.6417 & 0.7335 & 0.7335 & 129 & 0.6390 & 0.7003 & 0.7003\\
\hline
6 & 0.7762 & 0.8851 & 0.9066 & 56 & 0.6530 & 0.7328 & 0.7328 & 130 & 0.6440 & 0.7001 & 0.7001\\
\hline
7 & 0.6789 & 0.8769 & 0.8902 & 57 & 0.6418 & 0.7320 & 0.7320 & 139 & 0.6385 & 0.6976 & 0.6976\\
\hline
8 & 0.7435 & 0.8681 & 0.8766 & 58 & 0.6520 & 0.7308 & 0.7308 & 140 & 0.6429 & 0.6974 & 0.6974\\
\hline
9 & 0.6691 & 0.8585 & 0.8646 & 59 & 0.6410 & 0.7299 & 0.7299 & 149 & 0.6385 & 0.6956 & 0.6956\\
\hline
10 & 0.7229 & 0.8498 & 0.8544 & 60 & 0.6515 & 0.7292 & 0.7292 & 150 & 0.6430 & 0.6958 & 0.6958\\
\hline
11 & 0.6626 & 0.8419 & 0.8454 & 61 & 0.6407 & 0.7283 & 0.7283 & 159 & 0.6380 & 0.6934 & 0.6934\\
\hline
12 & 0.7091 & 0.8345 & 0.8372 & 62 & 0.6515 & 0.7282 & 0.7282 & 160 & 0.6423 & 0.6936 & 0.6936\\
\hline
13 & 0.6589 & 0.8284 & 0.8303 & 63 & 0.6408 & 0.7270 & 0.7270 & 169 & 0.6385 & 0.6923 & 0.6923\\
\hline
14 & 0.6989 & 0.8220 & 0.8233 & 64 & 0.6505 & 0.7263 & 0.7263 & 170 & 0.6418 & 0.6918 & 0.6918\\
\hline
15 & 0.6558 & 0.8168 & 0.8177 & 65 & 0.6406 & 0.7255 & 0.7255 & 179 & 0.6381 & 0.6904 & 0.6904\\
\hline
16 & 0.6918 & 0.8121 & 0.8127 & 66 & 0.6502 & 0.7249 & 0.7249 & 180 & 0.6416 & 0.6903 & 0.6903\\
\hline
17 & 0.6535 & 0.8070 & 0.8075 & 67 & 0.6408 & 0.7245 & 0.7245 & 189 & 0.6380 & 0.6890 & 0.6890\\
\hline
18 & 0.6858 & 0.8027 & 0.8030 & 68 & 0.6499 & 0.7238 & 0.7238 & 190 & 0.6418 & 0.6893 & 0.6893\\
\hline
19 & 0.6515 & 0.7983 & 0.7986 & 69 & 0.6405 & 0.7233 & 0.7233 & 199 & 0.6383 & 0.6880 & 0.6880\\
\hline
20 & 0.6811 & 0.7948 & 0.7950 & 70 & 0.6491 & 0.7221 & 0.7221 & 200 & 0.6415 & 0.6881 & 0.6881\\
\hline
21 & 0.6503 & 0.7912 & 0.7914 & 71 & 0.6404 & 0.7218 & 0.7218 & 249 & 0.6378 & 0.6824 & 0.6824\\
\hline
22 & 0.6772 & 0.7881 & 0.7882 & 72 & 0.6492 & 0.7215 & 0.7215 & 250 & 0.6403 & 0.6823 & 0.6823\\
\hline
23 & 0.6488 & 0.7842 & 0.7843 & 73 & 0.6403 & 0.7208 & 0.7208 & 299 & 0.6374 & 0.6782 & 0.6782\\
\hline
24 & 0.6737 & 0.7816 & 0.7816 & 74 & 0.6491 & 0.7202 & 0.7202 & 300 & 0.6395 & 0.6781 & 0.6781\\
\hline
25 & 0.6480 & 0.7789 & 0.7790 & 75 & 0.6402 & 0.7197 & 0.7197 & 349 & 0.6377 & 0.6756 & 0.6756\\
\hline
26 & 0.6709 & 0.7763 & 0.7763 & 76 & 0.6487 & 0.7194 & 0.7194 & 350 & 0.6393 & 0.6752 & 0.6752\\
\hline
27 & 0.6466 & 0.7733 & 0.7733 & 77 & 0.6402 & 0.7185 & 0.7185 & 399 & 0.6372 & 0.6726 & 0.6726\\
\hline
28 & 0.6687 & 0.7714 & 0.7714 & 78 & 0.6481 & 0.7181 & 0.7181 & 400 & 0.6385 & 0.6723 & 0.6723\\
\hline
29 & 0.6467 & 0.7692 & 0.7692 & 79 & 0.6401 & 0.7175 & 0.7175 & 449 & 0.6371 & 0.6705 & 0.6705\\
\hline
30 & 0.6663 & 0.7667 & 0.7667 & 80 & 0.6479 & 0.7172 & 0.7172 & 450 & 0.6386 & 0.6707 & 0.6707\\
\hline
31 & 0.6457 & 0.7650 & 0.7650 & 81 & 0.6399 & 0.7164 & 0.7164 & 499 & 0.6372 & 0.6689 & 0.6689\\
\hline
32 & 0.6645 & 0.7627 & 0.7627 & 82 & 0.6477 & 0.7162 & 0.7162 & 500 & 0.6381 & 0.6685 & 0.6685\\
\hline
33 & 0.6448 & 0.7606 & 0.7606 & 83 & 0.6398 & 0.7155 & 0.7155 & 600 & 0.6383 & 0.6662 & 0.6662\\
\hline
34 & 0.6626 & 0.7590 & 0.7590 & 84 & 0.6475 & 0.7152 & 0.7152 & 700 & 0.6375 & 0.6635 & 0.6635\\
\hline
35 & 0.6445 & 0.7571 & 0.7571 & 85 & 0.6401 & 0.7147 & 0.7147 & 800 & 0.6378 & 0.6622 & 0.6622\\
\hline
36 & 0.6614 & 0.7557 & 0.7557 & 86 & 0.6469 & 0.7140 & 0.7140 & 900 & 0.6380 & 0.6610 & 0.6610\\
\hline
37 & 0.6439 & 0.7541 & 0.7541 & 87 & 0.6401 & 0.7141 & 0.7141 & 1000 & 0.6375 & 0.6593 & 0.6593\\
\hline
38 & 0.6598 & 0.7523 & 0.7523 & 88 & 0.6470 & 0.7134 & 0.7134 & 1500 & 0.6371 & 0.6551 & 0.6551\\
\hline
39 & 0.6439 & 0.7516 & 0.7516 & 89 & 0.6397 & 0.7129 & 0.7129 & 2000 & 0.6372 & 0.6528 & 0.6528\\
\hline
40 & 0.6589 & 0.7498 & 0.7498 & 90 & 0.6468 & 0.7127 & 0.7127 & 3000 & 0.6365 & 0.6493 & 0.6493\\
\hline
41 & 0.6434 & 0.7485 & 0.7485 & 91 & 0.6394 & 0.7118 & 0.7118 & 4000 & 0.6369 & 0.6481 & 0.6481\\
\hline
42 & 0.6586 & 0.7477 & 0.7477 & 92 & 0.6469 & 0.7122 & 0.7122 & 5000 & 0.6366 & 0.6466 & 0.6466\\
\hline
43 & 0.6429 & 0.7459 & 0.7459 & 93 & 0.6401 & 0.7117 & 0.7117 & 6000 & 0.6365 & 0.6457 & 0.6457\\
\hline
44 & 0.6571 & 0.7447 & 0.7447 & 94 & 0.6459 & 0.7105 & 0.7105 & 7000 & 0.6367 & 0.6451 & 0.6451\\
\hline
45 & 0.6425 & 0.7433 & 0.7433 & 95 & 0.6393 & 0.7104 & 0.7104 & 8000 & 0.6369 & 0.6448 & 0.6448\\
\hline
46 & 0.6563 & 0.7423 & 0.7423 & 96 & 0.6459 & 0.7101 & 0.7101 & 9000 & 0.6369 & 0.6444 & 0.6444\\
\hline
47 & 0.6421 & 0.7410 & 0.7410 & 97 & 0.6393 & 0.7096 & 0.7096 & 10000 & 0.6370 & 0.6440 & 0.6440\\
\hline
48 & 0.6556 & 0.7405 & 0.7405 & 98 & 0.6463 & 0.7098 & 0.7098 & 25000 & 0.6367 & 0.6413 & 0.6413\\
\hline
49 & 0.6425 & 0.7393 & 0.7393 & 99 & 0.6394 & 0.7090 & 0.7090 & 50000 & 0.6366 & 0.6398 & 0.6398\\
\hline
50 & 0.6541 & 0.7377 & 0.7377 & 100 & 0.6457 & 0.7087 & 0.7087 & 100000 & 0.6367 & 0.6390 & 0.6390\\
\hline
\end{tabular}
\end{table}

\clearpage

\clearpage

\hypertarget{sec:summary}{%
\section{Summary}\label{sec:summary}}

In this paper, we have introduced
the \emph{quantile absolute deviation around the median} (\(\QAD\)) which is
a generalization of the \emph{median absolute deviation around the median} (\(\MAD\)).
For a finite sample \(X\) of size \(n\) it is defined as follows:

\[
\QAD_n(X, p) = K_{p,n} \Q(|X - \median(X)|, p),
\]

where \(K_{p,n}\) is a consistency constant.
The parameter \(p\) allows us not only to get a broad perspective on the dispersion behavior in the non-parametric case,
but also customize the trade-off between the breakdown point (which is \(1-p\)) and the Gaussian efficiency.
In addition to the \(\MAD_n(X)\) which is \(\QAD_n(X, 0.5)\), we have introduced two rule of thumbs for choosing \(p\).
More specifically, we have considered the standard \(\QAD\) (\(\SQAD\)) and the optimal \(\QAD\) (\(\OQAD\)):

\[
\begin{split}
\SQAD(X) = \QAD(X, \rho_s), \quad \rho_s \approx 0.6827,\\
\OQAD(X) = \QAD(X, \rho_o), \quad \rho_o \approx 0.8617.
\end{split}
\]

The asymptotic properties of the \(\MAD\), the \(\SQAD\), and the \(\OQAD\) are presented in Table~\ref{tab:summary}.

\begin{longtable}[]{@{}
  >{\raggedleft\arraybackslash}p{(\columnwidth - 6\tabcolsep) * \real{0.1549}}
  >{\raggedleft\arraybackslash}p{(\columnwidth - 6\tabcolsep) * \real{0.2394}}
  >{\raggedleft\arraybackslash}p{(\columnwidth - 6\tabcolsep) * \real{0.2958}}
  >{\raggedleft\arraybackslash}p{(\columnwidth - 6\tabcolsep) * \real{0.3099}}@{}}
\caption{\label{tab:summary} Approximated asymptotic properties of considered dispersion estimator.}\tabularnewline
\toprule()
\begin{minipage}[b]{\linewidth}\raggedleft
Estimator
\end{minipage} & \begin{minipage}[b]{\linewidth}\raggedleft
Breakdown point
\end{minipage} & \begin{minipage}[b]{\linewidth}\raggedleft
Gaussian efficiency
\end{minipage} & \begin{minipage}[b]{\linewidth}\raggedleft
Consistency constant
\end{minipage} \\
\midrule()
\endfirsthead
\toprule()
\begin{minipage}[b]{\linewidth}\raggedleft
Estimator
\end{minipage} & \begin{minipage}[b]{\linewidth}\raggedleft
Breakdown point
\end{minipage} & \begin{minipage}[b]{\linewidth}\raggedleft
Gaussian efficiency
\end{minipage} & \begin{minipage}[b]{\linewidth}\raggedleft
Consistency constant
\end{minipage} \\
\midrule()
\endhead
\(\MAD\) & 50.00\% & 36.75\% & 1.482602 \\
\(\SQAD\) & 31.73\% & 54.06\% & 1.000000 \\
\(\OQAD\) & 13.83\% & 65.22\% & 0.674731 \\
\bottomrule()
\end{longtable}

We also suggest using the trimmed Harrell-Davis median estimator \(\THDME_p\)
as a complimentary median estimator for \(\QAD(X, p)\) since it has the same asymptotic breakdown point.
Various \(\QAD(X, p)\) dispersion estimators including the \(\SQAD\) and the \(\OQAD\)
can be reasonable replacements for the \(\MAD\) when the breakdown point of \(50\%\) is not required.

\hypertarget{disclosure-statement}{%
\section*{Disclosure statement}\label{disclosure-statement}}
\addcontentsline{toc}{section}{Disclosure statement}

The author reports there are no competing interests to declare.

\hypertarget{data-and-source-code-availability}{%
\section*{Data and source code availability}\label{data-and-source-code-availability}}
\addcontentsline{toc}{section}{Data and source code availability}

The source code of this paper, the source code of all simulations,
and the simulation results are available on GitHub:
\url{https://github.com/AndreyAkinshin/paper-qad}.

\hypertarget{acknowledgments}{%
\section*{Acknowledgments}\label{acknowledgments}}
\addcontentsline{toc}{section}{Acknowledgments}

The author thanks Ivan Pashchenko for valuable discussions.

\clearpage

\hypertarget{appendix-appendix}{%
\appendix}

\hypertarget{sec:refimpl}{%
\section{Reference implementation}\label{sec:refimpl}}

Here is an R implementation of the \(\QAD\), the \(\SQAD\), and the \(\OQAD\):

\begin{Shaded}
\begin{Highlighting}[]
\CommentTok{\# Quantile absolute deviation}
\NormalTok{qad }\OtherTok{\textless{}{-}} \ControlFlowTok{function}\NormalTok{(x, p, }\AttributeTok{constant =} \DecValTok{1}\NormalTok{) \{}
\NormalTok{  x }\OtherTok{\textless{}{-}}\NormalTok{ x[}\SpecialCharTok{!}\FunctionTok{is.na}\NormalTok{(x)]}
  \FunctionTok{as.numeric}\NormalTok{(}\FunctionTok{quantile}\NormalTok{(}\FunctionTok{abs}\NormalTok{(x }\SpecialCharTok{{-}} \FunctionTok{median}\NormalTok{(x)), p)) }\SpecialCharTok{*}\NormalTok{ constant}
\NormalTok{\}}

\CommentTok{\# Standard quantile absolute deviation}
\NormalTok{sqad }\OtherTok{\textless{}{-}} \ControlFlowTok{function}\NormalTok{(x, }\AttributeTok{constant =} \ConstantTok{NULL}\NormalTok{) \{}
\NormalTok{  x }\OtherTok{\textless{}{-}}\NormalTok{ x[}\SpecialCharTok{!}\FunctionTok{is.na}\NormalTok{(x)]}
\NormalTok{  n }\OtherTok{\textless{}{-}} \FunctionTok{length}\NormalTok{(x)}
  \ControlFlowTok{if}\NormalTok{ (n }\SpecialCharTok{==} \DecValTok{0}\NormalTok{) }\FunctionTok{return}\NormalTok{(}\ConstantTok{NA}\NormalTok{)}
\NormalTok{  constants }\OtherTok{\textless{}{-}}
    \FunctionTok{c}\NormalTok{(    }\ConstantTok{NA}\NormalTok{, }\FloatTok{1.7724}\NormalTok{, }\FloatTok{1.3506}\NormalTok{, }\FloatTok{1.3762}\NormalTok{, }\FloatTok{1.1881}\NormalTok{, }\FloatTok{1.1773}\NormalTok{, }\FloatTok{1.1289}\NormalTok{, }\FloatTok{1.1248}\NormalTok{, }\FloatTok{1.0920}\NormalTok{, }\FloatTok{1.0943}\NormalTok{,}
      \FloatTok{1.0764}\NormalTok{, }\FloatTok{1.0738}\NormalTok{, }\FloatTok{1.0630}\NormalTok{, }\FloatTok{1.0637}\NormalTok{, }\FloatTok{1.0533}\NormalTok{, }\FloatTok{1.0537}\NormalTok{, }\FloatTok{1.0482}\NormalTok{, }\FloatTok{1.0468}\NormalTok{, }\FloatTok{1.0419}\NormalTok{, }\FloatTok{1.0429}\NormalTok{,}
      \FloatTok{1.0377}\NormalTok{, }\FloatTok{1.0376}\NormalTok{, }\FloatTok{1.0351}\NormalTok{, }\FloatTok{1.0343}\NormalTok{, }\FloatTok{1.0314}\NormalTok{, }\FloatTok{1.0320}\NormalTok{, }\FloatTok{1.0292}\NormalTok{, }\FloatTok{1.0290}\NormalTok{, }\FloatTok{1.0272}\NormalTok{, }\FloatTok{1.0271}\NormalTok{,}
      \FloatTok{1.0251}\NormalTok{, }\FloatTok{1.0253}\NormalTok{, }\FloatTok{1.0238}\NormalTok{, }\FloatTok{1.0235}\NormalTok{, }\FloatTok{1.0223}\NormalTok{, }\FloatTok{1.0224}\NormalTok{, }\FloatTok{1.0210}\NormalTok{, }\FloatTok{1.0210}\NormalTok{, }\FloatTok{1.0201}\NormalTok{, }\FloatTok{1.0199}\NormalTok{,}
      \FloatTok{1.0189}\NormalTok{, }\FloatTok{1.0192}\NormalTok{, }\FloatTok{1.0180}\NormalTok{, }\FloatTok{1.0180}\NormalTok{, }\FloatTok{1.0174}\NormalTok{, }\FloatTok{1.0172}\NormalTok{, }\FloatTok{1.0165}\NormalTok{, }\FloatTok{1.0166}\NormalTok{, }\FloatTok{1.0158}\NormalTok{, }\FloatTok{1.0158}\NormalTok{,}
      \FloatTok{1.0152}\NormalTok{, }\FloatTok{1.0152}\NormalTok{, }\FloatTok{1.0146}\NormalTok{, }\FloatTok{1.0146}\NormalTok{, }\FloatTok{1.0141}\NormalTok{, }\FloatTok{1.0140}\NormalTok{, }\FloatTok{1.0135}\NormalTok{, }\FloatTok{1.0137}\NormalTok{, }\FloatTok{1.0130}\NormalTok{, }\FloatTok{1.0131}\NormalTok{,}
      \FloatTok{1.0127}\NormalTok{, }\FloatTok{1.0126}\NormalTok{, }\FloatTok{1.0123}\NormalTok{, }\FloatTok{1.0124}\NormalTok{, }\FloatTok{1.0118}\NormalTok{, }\FloatTok{1.0119}\NormalTok{, }\FloatTok{1.0115}\NormalTok{, }\FloatTok{1.0115}\NormalTok{, }\FloatTok{1.0111}\NormalTok{, }\FloatTok{1.0112}\NormalTok{,}
      \FloatTok{1.0108}\NormalTok{, }\FloatTok{1.0108}\NormalTok{, }\FloatTok{1.0106}\NormalTok{, }\FloatTok{1.0106}\NormalTok{, }\FloatTok{1.0102}\NormalTok{, }\FloatTok{1.0103}\NormalTok{, }\FloatTok{1.0100}\NormalTok{, }\FloatTok{1.0100}\NormalTok{, }\FloatTok{1.0097}\NormalTok{, }\FloatTok{1.0097}\NormalTok{,}
      \FloatTok{1.0095}\NormalTok{, }\FloatTok{1.0095}\NormalTok{, }\FloatTok{1.0093}\NormalTok{, }\FloatTok{1.0092}\NormalTok{, }\FloatTok{1.0090}\NormalTok{, }\FloatTok{1.0091}\NormalTok{, }\FloatTok{1.0089}\NormalTok{, }\FloatTok{1.0088}\NormalTok{, }\FloatTok{1.0086}\NormalTok{, }\FloatTok{1.0086}\NormalTok{,}
      \FloatTok{1.0084}\NormalTok{, }\FloatTok{1.0084}\NormalTok{, }\FloatTok{1.0082}\NormalTok{, }\FloatTok{1.0082}\NormalTok{, }\FloatTok{1.0081}\NormalTok{, }\FloatTok{1.0081}\NormalTok{, }\FloatTok{1.0079}\NormalTok{, }\FloatTok{1.0079}\NormalTok{, }\FloatTok{1.0078}\NormalTok{, }\FloatTok{1.0077}\NormalTok{)}
  \ControlFlowTok{if}\NormalTok{ (}\FunctionTok{is.null}\NormalTok{(constant)) }
\NormalTok{    constant }\OtherTok{\textless{}{-}} \ControlFlowTok{if}\NormalTok{ (n }\SpecialCharTok{\textless{}=} \DecValTok{100}\NormalTok{) constants[n] }\ControlFlowTok{else} \DecValTok{1} \SpecialCharTok{+} \FloatTok{0.762} \SpecialCharTok{/}\NormalTok{ n }\SpecialCharTok{+} \FloatTok{0.967} \SpecialCharTok{/}\NormalTok{ n}\SpecialCharTok{\^{}}\DecValTok{2}
  \FunctionTok{qad}\NormalTok{(x, }\FunctionTok{pnorm}\NormalTok{(}\DecValTok{1}\NormalTok{) }\SpecialCharTok{{-}} \FunctionTok{pnorm}\NormalTok{(}\SpecialCharTok{{-}}\DecValTok{1}\NormalTok{), constant)}
\NormalTok{\}}

\CommentTok{\# Optimal quantile absolute deviation}
\NormalTok{oqad }\OtherTok{\textless{}{-}} \ControlFlowTok{function}\NormalTok{(x, }\AttributeTok{constant =} \ConstantTok{NULL}\NormalTok{) \{}
\NormalTok{  x }\OtherTok{\textless{}{-}}\NormalTok{ x[}\SpecialCharTok{!}\FunctionTok{is.na}\NormalTok{(x)]}
\NormalTok{  n }\OtherTok{\textless{}{-}} \FunctionTok{length}\NormalTok{(x)}
  \ControlFlowTok{if}\NormalTok{ (n }\SpecialCharTok{==} \DecValTok{0}\NormalTok{) }\FunctionTok{return}\NormalTok{(}\ConstantTok{NA}\NormalTok{)}
\NormalTok{  constants }\OtherTok{\textless{}{-}}
    \FunctionTok{c}\NormalTok{(    }\ConstantTok{NA}\NormalTok{, }\FloatTok{1.7729}\NormalTok{, }\FloatTok{0.9788}\NormalTok{, }\FloatTok{0.9205}\NormalTok{, }\FloatTok{0.8194}\NormalTok{, }\FloatTok{0.8110}\NormalTok{, }\FloatTok{0.7792}\NormalTok{, }\FloatTok{0.7828}\NormalTok{, }\FloatTok{0.7600}\NormalTok{, }\FloatTok{0.7535}\NormalTok{,}
      \FloatTok{0.7388}\NormalTok{, }\FloatTok{0.7365}\NormalTok{, }\FloatTok{0.7282}\NormalTok{, }\FloatTok{0.7284}\NormalTok{, }\FloatTok{0.7241}\NormalTok{, }\FloatTok{0.7234}\NormalTok{, }\FloatTok{0.7170}\NormalTok{, }\FloatTok{0.7155}\NormalTok{, }\FloatTok{0.7113}\NormalTok{, }\FloatTok{0.7110}\NormalTok{,}
      \FloatTok{0.7083}\NormalTok{, }\FloatTok{0.7088}\NormalTok{, }\FloatTok{0.7068}\NormalTok{, }\FloatTok{0.7056}\NormalTok{, }\FloatTok{0.7030}\NormalTok{, }\FloatTok{0.7024}\NormalTok{, }\FloatTok{0.7006}\NormalTok{, }\FloatTok{0.7006}\NormalTok{, }\FloatTok{0.6995}\NormalTok{, }\FloatTok{0.6998}\NormalTok{,}
      \FloatTok{0.6979}\NormalTok{, }\FloatTok{0.6974}\NormalTok{, }\FloatTok{0.6960}\NormalTok{, }\FloatTok{0.6958}\NormalTok{, }\FloatTok{0.6949}\NormalTok{, }\FloatTok{0.6949}\NormalTok{, }\FloatTok{0.6944}\NormalTok{, }\FloatTok{0.6940}\NormalTok{, }\FloatTok{0.6929}\NormalTok{, }\FloatTok{0.6927}\NormalTok{,}
      \FloatTok{0.6918}\NormalTok{, }\FloatTok{0.6918}\NormalTok{, }\FloatTok{0.6913}\NormalTok{, }\FloatTok{0.6914}\NormalTok{, }\FloatTok{0.6907}\NormalTok{, }\FloatTok{0.6904}\NormalTok{, }\FloatTok{0.6897}\NormalTok{, }\FloatTok{0.6896}\NormalTok{, }\FloatTok{0.6891}\NormalTok{, }\FloatTok{0.6892}\NormalTok{,}
      \FloatTok{0.6888}\NormalTok{, }\FloatTok{0.6887}\NormalTok{, }\FloatTok{0.6882}\NormalTok{, }\FloatTok{0.6880}\NormalTok{, }\FloatTok{0.6875}\NormalTok{, }\FloatTok{0.6875}\NormalTok{, }\FloatTok{0.6871}\NormalTok{, }\FloatTok{0.6872}\NormalTok{, }\FloatTok{0.6870}\NormalTok{, }\FloatTok{0.6868}\NormalTok{,}
      \FloatTok{0.6863}\NormalTok{, }\FloatTok{0.6862}\NormalTok{, }\FloatTok{0.6859}\NormalTok{, }\FloatTok{0.6859}\NormalTok{, }\FloatTok{0.6857}\NormalTok{, }\FloatTok{0.6858}\NormalTok{, }\FloatTok{0.6854}\NormalTok{, }\FloatTok{0.6853}\NormalTok{, }\FloatTok{0.6850}\NormalTok{, }\FloatTok{0.6849}\NormalTok{,}
      \FloatTok{0.6847}\NormalTok{, }\FloatTok{0.6847}\NormalTok{, }\FloatTok{0.6846}\NormalTok{, }\FloatTok{0.6845}\NormalTok{, }\FloatTok{0.6842}\NormalTok{, }\FloatTok{0.6841}\NormalTok{, }\FloatTok{0.6839}\NormalTok{, }\FloatTok{0.6839}\NormalTok{, }\FloatTok{0.6837}\NormalTok{, }\FloatTok{0.6838}\NormalTok{,}
      \FloatTok{0.6836}\NormalTok{, }\FloatTok{0.6834}\NormalTok{, }\FloatTok{0.6833}\NormalTok{, }\FloatTok{0.6832}\NormalTok{, }\FloatTok{0.6831}\NormalTok{, }\FloatTok{0.6830}\NormalTok{, }\FloatTok{0.6829}\NormalTok{, }\FloatTok{0.6830}\NormalTok{, }\FloatTok{0.6827}\NormalTok{, }\FloatTok{0.6827}\NormalTok{,}
      \FloatTok{0.6825}\NormalTok{, }\FloatTok{0.6825}\NormalTok{, }\FloatTok{0.6823}\NormalTok{, }\FloatTok{0.6823}\NormalTok{, }\FloatTok{0.6823}\NormalTok{, }\FloatTok{0.6822}\NormalTok{, }\FloatTok{0.6820}\NormalTok{, }\FloatTok{0.6820}\NormalTok{, }\FloatTok{0.6819}\NormalTok{, }\FloatTok{0.6819}\NormalTok{)}
  \ControlFlowTok{if}\NormalTok{ (}\FunctionTok{is.null}\NormalTok{(constant)) }
\NormalTok{    constant }\OtherTok{\textless{}{-}} \ControlFlowTok{if}\NormalTok{ (n }\SpecialCharTok{\textless{}=} \DecValTok{100}\NormalTok{) constants[n] }\ControlFlowTok{else} \FloatTok{0.6747309} \SpecialCharTok{*}\NormalTok{ (}\DecValTok{1} \SpecialCharTok{+} \FloatTok{1.047} \SpecialCharTok{/}\NormalTok{ n }\SpecialCharTok{+} \FloatTok{1.193} \SpecialCharTok{/}\NormalTok{ n}\SpecialCharTok{\^{}}\DecValTok{2}\NormalTok{)}
  \FunctionTok{qad}\NormalTok{(x, }\FloatTok{0.861678977787423}\NormalTok{, constant)}
\NormalTok{\}}
\end{Highlighting}
\end{Shaded}

\clearpage

And here is an R implementation of the \(\THDME\), the \(\STHDME\), and the \(\OTHDME\):

\begin{Shaded}
\begin{Highlighting}[]
\CommentTok{\# Trimmed Harrell{-}Davis median estimator based on the highest density interval of width D}
\NormalTok{thdme }\OtherTok{\textless{}{-}} \ControlFlowTok{function}\NormalTok{(x, D) \{}
\NormalTok{  x }\OtherTok{\textless{}{-}}\NormalTok{ x[}\SpecialCharTok{!}\FunctionTok{is.na}\NormalTok{(x)]}
\NormalTok{  n }\OtherTok{\textless{}{-}} \FunctionTok{length}\NormalTok{(x)}
  \ControlFlowTok{if}\NormalTok{ (n }\SpecialCharTok{==} \DecValTok{0}\NormalTok{) }\FunctionTok{return}\NormalTok{(}\ConstantTok{NA}\NormalTok{)}
  \ControlFlowTok{if}\NormalTok{ (n }\SpecialCharTok{==} \DecValTok{1}\NormalTok{) }\FunctionTok{return}\NormalTok{(x)}
\NormalTok{  x }\OtherTok{\textless{}{-}} \FunctionTok{sort}\NormalTok{(x)}
\NormalTok{  a }\OtherTok{\textless{}{-}}\NormalTok{ (n }\SpecialCharTok{+} \DecValTok{1}\NormalTok{) }\SpecialCharTok{/} \DecValTok{2}\NormalTok{; b }\OtherTok{\textless{}{-}}\NormalTok{ (n }\SpecialCharTok{+} \DecValTok{1}\NormalTok{) }\SpecialCharTok{/} \DecValTok{2}
\NormalTok{  hdi }\OtherTok{\textless{}{-}} \FunctionTok{c}\NormalTok{(}\FloatTok{0.5} \SpecialCharTok{{-}}\NormalTok{ D }\SpecialCharTok{/} \DecValTok{2}\NormalTok{, }\FloatTok{0.5} \SpecialCharTok{+}\NormalTok{ D }\SpecialCharTok{/} \DecValTok{2}\NormalTok{)}
\NormalTok{  hdiCdf }\OtherTok{\textless{}{-}} \FunctionTok{pbeta}\NormalTok{(hdi, a, b)}
\NormalTok{  cdf }\OtherTok{\textless{}{-}} \ControlFlowTok{function}\NormalTok{(xs) \{}
\NormalTok{    xs[xs }\SpecialCharTok{\textless{}=}\NormalTok{ hdi[}\DecValTok{1}\NormalTok{]] }\OtherTok{\textless{}{-}}\NormalTok{ hdi[}\DecValTok{1}\NormalTok{]}
\NormalTok{    xs[xs }\SpecialCharTok{\textgreater{}=}\NormalTok{ hdi[}\DecValTok{2}\NormalTok{]] }\OtherTok{\textless{}{-}}\NormalTok{ hdi[}\DecValTok{2}\NormalTok{]}
\NormalTok{    (}\FunctionTok{pbeta}\NormalTok{(xs, a, b) }\SpecialCharTok{{-}}\NormalTok{ hdiCdf[}\DecValTok{1}\NormalTok{]) }\SpecialCharTok{/}\NormalTok{ (hdiCdf[}\DecValTok{2}\NormalTok{] }\SpecialCharTok{{-}}\NormalTok{ hdiCdf[}\DecValTok{1}\NormalTok{])}
\NormalTok{  \}}
\NormalTok{  iL }\OtherTok{\textless{}{-}} \FunctionTok{floor}\NormalTok{(hdi[}\DecValTok{1}\NormalTok{] }\SpecialCharTok{*}\NormalTok{ n); iR }\OtherTok{\textless{}{-}} \FunctionTok{ceiling}\NormalTok{(hdi[}\DecValTok{2}\NormalTok{] }\SpecialCharTok{*}\NormalTok{ n)}
\NormalTok{  cdfs }\OtherTok{\textless{}{-}} \FunctionTok{cdf}\NormalTok{(iL}\SpecialCharTok{:}\NormalTok{iR}\SpecialCharTok{/}\NormalTok{n)}
\NormalTok{  W }\OtherTok{\textless{}{-}} \FunctionTok{tail}\NormalTok{(cdfs, }\SpecialCharTok{{-}}\DecValTok{1}\NormalTok{) }\SpecialCharTok{{-}} \FunctionTok{head}\NormalTok{(cdfs, }\SpecialCharTok{{-}}\DecValTok{1}\NormalTok{)}
  \FunctionTok{sum}\NormalTok{(x[(iL }\SpecialCharTok{+} \DecValTok{1}\NormalTok{)}\SpecialCharTok{:}\NormalTok{iR] }\SpecialCharTok{*}\NormalTok{ W)}
\NormalTok{\}}

\CommentTok{\# Standard trimmed Harrell{-}Davis median estimator}
\NormalTok{sthdme }\OtherTok{\textless{}{-}} \ControlFlowTok{function}\NormalTok{(x) }\FunctionTok{thdme}\NormalTok{(x, }\FunctionTok{pnorm}\NormalTok{(}\DecValTok{1}\NormalTok{) }\SpecialCharTok{{-}} \FunctionTok{pnorm}\NormalTok{(}\SpecialCharTok{{-}}\DecValTok{1}\NormalTok{))}

\CommentTok{\# Optimal trimmed Harrell{-}Davis median estimator}
\NormalTok{othdme }\OtherTok{\textless{}{-}} \ControlFlowTok{function}\NormalTok{(x) }\FunctionTok{thdme}\NormalTok{(x, }\FloatTok{0.861678977787423}\NormalTok{)}
\end{Highlighting}
\end{Shaded}

\clearpage

\printbibliography[title=References]

\end{document}